\documentclass[twocolumn]{aastex631}

\begin{document}

\title{{Giant Eruptions in Massive Stars and their Effect on the Stellar Structure}}

\shorttitle{Giant Eruptions in Massive Stars}
\shortauthors{B. Mukhija and A. Kashi}

\author[0009-0007-1450-6490]{Bhawna Mukhija}
\affiliation{Department of Physics, Ariel University, Ariel, 4070000, Israel}
\email{bhawnam@ariel.ac.il}

\author[0000-0002-7840-0181]{Amit Kashi}
\email{kashi@ariel.ac.il}
\affiliation{Department of Physics, Ariel University, Ariel, 4070000, Israel}
\affiliation{Astrophysics, Geophysics, and Space Science (AGASS) Center, Ariel University, Ariel, 4070000, Israel}

\begin{abstract}
Giant eruptions (GE) in Luminous blue variables (LBVs) are years to decades-long episodes of enhanced mass loss from the outer layers of the star during which the star undergoes major changes in its physical and observed properties.
We use the \textsc{mesa} stellar evolution code to model the evolution of a $70~M_{\odot}$ star that undergoes a GE. We let the star evolve to the termination of the main sequence (MS) and when it reaches $T\simeq 19\,400 $ K we emulate a GE by removing mass from its outer layers, at a rate of $0.15~ M_{\odot}~\rm yr^{-1}$ for 20 years. As mass is being lost, the star contracts and releases a substantial amount of gravitational energy.
The star undergoes an initial $\simeq 3$ days of expansion followed by years of contraction.
During that time the star tries to reach an equilibrium state and as a result of loss in gravitational energy, its luminosity drops about one order of magnitude. As the GE terminates, we let the star continue to evolve without any further mass loss and track its recovery as it regains its equilibrium by adjusting its internal structure.
After $\simeq 87$ years it reaches a state very close to the one where the GE was first initiated. We suggest that at this point another GE or a cycle of GEs may occur.
\end{abstract}


\section{Introduction} \label{sec:intro}
Massive stars play a crucial role in understanding the formation and structure of the universe \citep[e.g.,][]{2012ASSL..384...43D, 2012ARA&A..50..107L, 2015ASSL..412...77V, 2016ApJ...817...66K,2022MNRAS.514.3736S,2022MNRAS.512.4116F}.
Because of their high mass, they are responsible for the nucleosynthesis of heavy elements in the universe, and they lead to the formation of core-collapse supernovae, neutron stars, and black holes \citep[e.g.,][]{2012ARA&A..50..107L,2013ascl.soft03015H, 2019Sci...363..474J, 2024ApJ...961...89V}.
The evolution of a star depends on various parameters, including its chemical composition, rotation speed, mass loss rate, and interactions with other stars in binary or multiple systems \citep[e.g.,][]{1986ARA&A..24..329C, 2000A&A...361..159M, 2012ARA&A..50..107L, 2017RSPTA.37560268S, 2022MNRAS.514.3736S, 2022MNRAS.516.3193K}. 
These stars possess powerful radiation driven stellar winds, which can remove a substantial amount of mass from the star's surface \citep[e.g.,][]{1982ApJ...259..282A,2000A&AS..141...23P, 2011A&A...528A..64S, 2012MNRAS.423L..92Q, 2012ApJ...751L..34V,2013A&A...560A...6G, 2016ApJ...821..109R, 2021A&A...648A..36B, 2022ARA&A..60..203V, 2022A&A...665A.133G, 2023Galax..11...68C}. These winds interact with the surrounding interstellar medium (ISM) and contribute to the chemical enrichment of the ISM and the formation of new stars \citep[e.g.,][]{2007MNRAS.374.1045H, 2014ApJ...785...82S, 2019ApJ...873..131G}.

Luminous Blue Variables (LBVs) are a rare and important class of massive stars that challenge their evolution and fate \citep[e.g.,][]{2012ASSL..384...43D}. They are known for their high mass loss rates $ \approx 10^{-6}$ ~to $~10^{-3}~ M_{\odot}~\rm yr^{-1}$ \citep{1997ASPC..120...58L, 2015ASSL..412...77V, 2020Galax...8...20W} and can undergo Giant eruptions (GE), during which their bolometric luminosity can increase for a few months to years, accompanied by rapid mass loss of even higher rates \citep{2016JPhCS.728b2008D, 2016ApJ...817...66K}. Eta Carinae, one of the most well-known examples of an LBV star \citep{1997ARA&A..35....1D}, experienced a giant eruption in the nineteenth century with a mass loss rate of up to $ \approx 1~ M_{\odot}~\rm yr^{-1}$. \citep[e.g.,][]{1999PASP..111.1124H, 2012Natur.486E...1D, 2012ASSL..384....1H, 2006MNRAS.372.1133G}. Another famous LBV star is P Cygni, which experienced dramatic eruption in the seventeenth century \citep[e.g.,][]{1988IrAJ...18..163D,1994PASP..106.1025H, 1997A&A...326.1117N,1999SSRv...90..493I, 1999ASPC..192...32D,2001ASPC..233..133N, 2010AJ....139.2269B, 2010MNRAS.405.1924K, 2012ASSL..384...43D,2018NewA...65...29M, 2020MNRAS.494..218R}. These eruptions can expel large amounts of stellar material into space, significantly affecting the star's evolution \citep{2016ApJ...817...66K,2017RSPTA.37560268S}. 

Giant LBV eruptions are sometimes referred to as supernova (SN) Impostors, a name that reflects the uncertainty in the nature of some of these events, leading to confusion with a real core-collapse SN \citep{2012ASSL..384.....D}.
One of the most enigmatic objects of this kind is SN 2009ip \citep[e.g.,][]{2010AJ....139.1451S, 2012ATel.4334....1D, 2013ApJ...767....1P, 2022MNRAS.515...71S}.
The star experienced a series of eruptions from 2009 to 2012 that were thought end in a terminal explosion \citep[e.g.,][]{2013MNRAS.430.1801M, 2022MNRAS.515...71S}, though other suggestions for a binary-induced GE were also proposed \citep{2013MNRAS.436.2484K}, that may be consistent with recent observations \citep{2023A&A...677L...1P}.
Among other impostors that show multiple eruptions and were associated with LBVs, we can list SN 2000ch \citep[e.g.,][]{2000IAUC.7415....1P, 2010MNRAS.408..181P, 2011MNRAS.415..773S,2012CBET.2976....1K, 2023MNRAS.521.1941A} that was also thought to be a bridging object between GEs and lower energy eruptions \citep{2013MNRAS.436.2484K}, and AT 2016blu \citep[e.g.,][]{2011MNRAS.415..773S,2012CBET.2976....1K,2023MNRAS.tmp.2585A} that showed not less than 19 outburst and might still be active.

As LBVs have very high luminosities \citep[e.g.,][]{2014ApJ...785...82S}, their envelope to approaches the Eddington limit, which in turn enhances their mass loss through stellar winds.
However, the dominance of radiation pressure over gas pressure can make the envelope unstable, and prone to rapid mass loss in the form of GEs.
These GEs show a sudden increase in luminosity, that can last for years or decades \citep[e.g.,][]{1994PASP..106.1025H, 1999PASP..111.1124H}. 
This rapid mass loss is often accompanied by the ejection of material in the form of shells or clumps, which can be observed as nebulous structures around the star \citep{2008MNRAS.389.1353V}.
Brightening of LBVs can also be a result of the S Doradus variability \citep[e.g.,][]{1994PASP..106.1025H, 2004ApJ...615..475S, 2012ApJ...751L..34V}, that may be related to the interaction of the wind and the helium opacity in the inflated envelope near the Eddington limit \citep{2021A&A...647A..99G}.

Several mechanisms were proposed for explaining LBV eruptions, among which we can list envelope instabilities \citep[e.g.,][]{1993MNRAS.263..375G, 2001ASPC..233..227G}, runaway mass loss  \citep[e.g.,][]{1993SSRv...66....7D, 1998A&A...329..551L, 2023IAUS..370..263S}, super-Eddington winds \citep[e.g.,]{2004ApJ...616..525O, Smith_2006} possibly involving porosity in the envelope \citep{2000ApJ...532L.137S}, and the critical rotation limit \citep{1998A&A...329..551L, 2012ARA&A..50..107L}. 
Other proposed mechanisms involve depositing a sizable quantity of extra energy deep inside a massive star \citep{2016ApJ...817...66K}. These mechanisms include unsteady burning \citep{2010MNRAS.405.2113D}, pulsational pair-instability \citep{RevModPhys.74.1015, 2007Natur.450..390W}, explosive shell burning instabilities \citep{2014ApJ...785...82S, 2010MNRAS.405.2113D} and wave-driven mass loss \citep[e.g.,][]{ 2000ARA&A..38..143M, smith2011explosions, 2012MNRAS.423L..92Q}.
Binary scenarios for LBV eruption include stellar collisions or mergers in binary systems \citep{2014ApJ...796..121J} and interaction in an eccentric orbit \citep[e.g.,][]{
2001MNRAS.325..584S, 2004ApJ...612.1060S,2010MNRAS.405.1924K, 2010ApJ...723..602K, 2020MNRAS.494.3186A}. 

One of the main questions yet to be studied with modern numerical tools is what occurs to the stellar structure of a massive star during a GE, and how such an event affects the future evolution of the star.
\citet{2009NewA...14..539H} simulated GE for a $ 190~M_{\odot} $ star, using the stellar evolution code \texttt{Evolve} \citep{1987ApJ...323..154H}. They removed mass at a rate of $1~{M_{\odot}}~{\rm yr^{-1}} $ from the outer envelope for 20 years, and found large fluctuations in luminosity and stellar radius.
They concluded that if a triggering mechanism starts removing the mass within a time much shorter than the thermal time scale of the star, a runway mass loss process will develop that will result in the removal of more mass. They further followed the star for another 200 years after the end of the GE to track its recovery, but due to code limitations, they were unable to obtain the correct values for the post-eruption phase.
\citet{2016ApJ...817...66K}, investigated the post-GE behavior of the VMS using two energy-conserving approaches. In the first method, they artificially removed outer layers of mass (a few $ M_{\odot}$) while concurrently diminishing the energy of the inner layers. The second method involved the synthetic transfer of energy from the core to the outer layers, resulting in the initiation of mass ejection. The primary objective was to analyze the recovery process after the eruption, and their findings led to the conclusion that the outflow was propelled by radial pulsations driven by the $ \kappa$ mechanism.
  
In the present paper, we attempt to explain the evolutionary and structural effect of GE in massive stars by creating an artificial eruption. We examine the changes in the stars' stellar characteristics and the evolutionary track on the HR diagram during the eruption phase. In addition, we let the star continue to evolve after the GE and analyze the stellar properties and evolutionary track during the recovery.

The paper is organized as follows. The basic assumptions and method of modeling the $ 70~M_{\odot} $ star are discussed in section \ref{2}.  The simulations and results are described in section \ref{3}. Our conclusion and summary are given in section \ref{4}.

\section{Physical ingredients of the model: \textsc{mesa} modeling}
\label{2}

We simulate a non-rotating $ 70~M_{\odot} $ star with metallicity $Z = 0.02 $ and introduce an episodic mass loss onto the outer envelope of the star. To perform this simulation, we utilize the 1-D stellar evolution code \textsc{mesa} (Modules for Experiments in Stellar Astrophysics; version r23.05.1), which has been referenced in various papers by Paxton \citep[e.g.,][]{2011ApJS..192....3P, 2013ApJS..208....4P, 2015ApJS..220...15P, 2018ApJS..234...34P, 2019ApJS..243...10P}. The model's parameters, such as the initial mass, metallicity, convection, overshooting, mixing, mass loss, and time step are discussed along with an explanation of their physical relevance below. We evolve a non-rotating star to isolate the impact of mass loss enhanced by the rotation on the evolution of the star \citep[e.g.,][]{1998A&A...329..551L, 2000ARA&A..38..143M, 2012ARA&A..50..107L, 2014A&A...564A..57M}. In \textsc{mesa}, the parameter for the initial mass fraction is defined as $ Y = Y_{\rm prim} + (\Delta Y / \Delta Z) / Z$. Here $ Y_{\rm prim}$ represents the initial He abundance, and $ Z$ defines the metallicity. While $ \Delta Y$ and $ \Delta Z$ represent the changes in the helium abundance and metallicity during the evolution.
The values, we used in our model are $ Y_{\rm prim}=0.24$ and $(\Delta  Y / \Delta  Z) / Z =2$. Both follow the default values in \textsc{mesa} \citep{10.1046/j.1365-8711.1998.01658.x}. We use the mixing length theory (\textsc{mlt}) given by \citet{1965ApJ...142..841H} to treat convection with mixing length parameter $\alpha_{\rm \textsc{mlt}}$ =1.6. \textsc{mesa} provides the opacity tables developed by \citet{1993ApJ...412..752I, 1996ApJ...464..943I}, featuring their OPAL type I opacity tables with fixed metal distributions as the default choice. The variations in the interior's abundance are sufficient to modify the opacity during the evolution. Thus,  it becomes necessary to use the OPAL type II opacity tables \citep{1996ApJ...464..943I}, which allow for time-dependent variation in the metal abundance in our model.

The \textsc{mesa} control defaults for the mass removal are given below in the section \ref{app}. The mass loss occurs in the outer layers of the star on a timescale ($ \approx 1000~\rm{s}$) that is shorter than the dynamic timescale of the star ($  t_{\rm dyn} = (2  R^{3}/{GM})^{1/2} $ $\approx 10^{4} ~\rm s$). 
\textsc{mesa} utilizes the atmosphere (\texttt{atm}) module to derive the surface temperature and surface pressure, which represents the conditions at the base of the stellar atmosphere. As the interior of the star changes, these values are used as boundary conditions. In our model, the (\texttt{atm}) option is set to the $T-\tau$, and we choose the Eddington grey approximation to define the temperature structure of the atmosphere.
We use the `Dutch' prescription to treat the hot winds of massive stars \citep{2009A&A...497..255G}. Additional parameters utilized in our simulation are listed in the section \ref{app} below.

When stellar evolution models use extrapolated standard O-star mass-loss rates ($ 10^{-10}$ to $10^{-5}~ M_{\odot}~ \rm yr^{-1}$), massive stars inflate to larger radii and low effective temperatures \citep{2000A&A...362..295V}. Thus, massive stars experience envelope inflation in addition to mass-loss rates, and stay to the cooler side of the HR diagram  \citep[e.g.,][]{1999PASJ...51..417I, 2006A&A...450..219P, 2012A&A...538A..40G, 2015A&A...580A..20S, 2015ApJ...813...74J}. 
Enhanced mass loss or \textsc{mlt++} approach in \textsc{mesa}, can be used to prevent this inflation \citep[e.g.,][]{2011ApJS..192....3P, 2021MNRAS.506.4473S, 2022MNRAS.514.3736S}. By using \textsc{mlt++}, the actual temperature gradients are reduced in super-adiabatic layers \citep{2022A&A...668A..90A} and cause the star to evolve to the hotter side of the HR diagram.

\section{Simulations and Results}
\label{3}

\begin{figure*}
  \includegraphics[trim= 0.0cm 0.0cm 1cm 1.0cm,clip=true,width=0.9\textwidth]{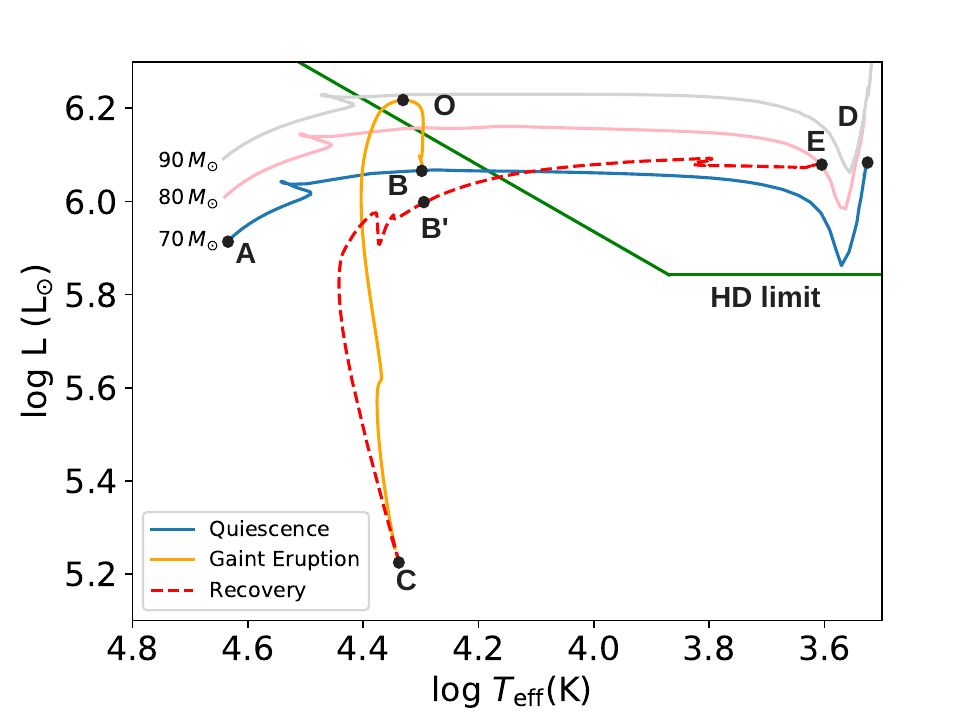} %trim=l b r t
  \centering
  
\caption{The track on the HR diagram of the simulations of a $70~ M_\odot$ star with metallicity $ Z=0.02$, starting at point A (zero-age main sequence). For the quiescence simulation (no GE) from point B to D (solid blue line) the track moves toward the cooler side of the HR diagram as the result of inflation. For the GE simulation, we start at point B with an enhanced mass loss rate of $0.15 ~{M_{\odot}}~{\rm yr^{-1}} $ for 20 years ( the track from point B through O to C along the orange line). In this phase, a transition occurs and the star evolves slightly towards the hotter side of the HR diagram. The track from point B to O shows the initial expansion and lasts only for $\rm \simeq 3.28~ days $. After that, the star tries to reach the equilibrium state and as a result of the mass loss and consequently loss of gravitational energy, there is a sharp drop in luminosity (track from points O to C). The track from point C to E represents the recovery phase over $ \simeq 3359~\rm years$.
Point B' is the closest point to the original state of the star before the eruption. The star reaches this point after $\simeq 87$ years. It is possible that more eruptions will occur next to this point or earlier.
In the background, we show for guidance the evolutionary tracks of $ 70~M_{\odot}$, $80~M_{\odot}$, and $ 90~M_{\odot}$ stars obtained by the \textsc{mesa} default directory \citep{2011ApJS..192....3P}.}
\label{hr}
\end{figure*}

\subsection{General considerations}
Consider a star with mass $M$, radius $ R$, and luminosity $  L$ just before the GE. The star Eddington luminosity is written as 
\begin{equation}
     L_{\rm Edd} = \frac{ 4 \pi GcM}{\kappa},
\end{equation}
where $ \kappa$ is the opacity, c is the speed of light, and $  G $ is the gravitational constant. In reality, some massive stars exceed their Eddington luminosity limit and are expected to be unstable \citep[e.g.,][]{2001ApJ...549.1093S, 2004ApJ...616..525O, 2004RMxAC..22..136V, 2009NewA...14..539H}, which is caused by the compression of the core region as mass loss begins on a time frame shorter than the dynamical time scale.
The triggering can result, for instance, from a tidal interaction with the companion that added energy to the envelope \citep{2001MNRAS.325..584S}.
To study this instability, we ran two models: the quiescence model where the star evolves without experiencing an eruption, and the GE model where we artificially remove mass from a $ 70~ M_{\odot} $ star and emulate the consequences of the eruption. We further examine the behavior of $ 70~M_{\odot} $ following these $\rm 20 $ years of a GE i.e., namely the recovery of the star after the GE.

\subsection{Quiescence }
\label{3.2}
In the quiescence model, we model an undisturbed $ 70~M_{\odot} $ star with no GE (blue line in Figure \ref{hr}).
The star evolves with only standard inlist file of  \textsc{MESA} from MS till the end of the He core burning.
Point B is not a special point for the quiescence model, but it marks when we start the GE simulation (see section \ref{3.3}) below.
Stellar parameters corresponding to the point B and D are mentioned in Table \ref{T1}. At point B, the star's age is  $ 2.918795~ \rm Myrs$ and it does not lose any mass till this point as we did not consider the Dutch wind lass loss due to the stellar winds in the early evolution. During the quiescence phase, the star undergoes radial inflation and moves toward the cooler side of the HR diagram. Thus, inflation is caused by the increased nuclear reaction energy in the shell. Since the envelope is unable to efficiently transport all the energy generated in the shell, it undergoes a cooling and expanding process. The cooling and expansion of the envelope are reflected in the increase of the stellar radii as mentioned in Table \ref{T1}. From point B to D, \textsc{mlt++} mixing and enhanced mass loss are turned off.

\begin{table*}
    \begin{tabular}{l | c c |c c c c c c c}
\hline
\hline
{} & \multicolumn{2}{c|}{Quiescence} & \multicolumn{4}{c}{Giant Eruption} \\
  \hline  
Stellar parameter & Point B & Point D & Point O & Point C & Point B'& Point E \\
\hline
\hline

     $\rm star~age~$($\rm M yrs$) &  2.918795 & 3.014429 & 2.91879579 & 2.918815 & 2.918902 & 2.9221.75 \\
     
     $ M ~( M_{\odot})$ & 70 & 70 &69.99 & 67 & 66.99 & 66.99\\

     $\log T_{\rm eff} ~(\rm K)$ & 4.29 & 3.53 & 4.31 & 4.35 & 4.30& 3.61 \\
     $L~ (L_{\odot})$ &  6.06      & 6.05  &6.21 & 5.28 & 5.99&6.07 \\
     $\log R ~ (R_{\odot})$  & 1.96 & 3.48 & 1.99 & 1.46 &1.90& 3.34  \\
     $\log  g ~ (\rm cm~s^{-2})$ & 2.35 & -0.69 &2.29 & 3.34& 2.44& -4.18\\ 
     $\log \dot{M}~(M_{\odot}~\rm yr^{-1})$      & - & -&  -0.82   & -0.82& - & - \\
     \hline \hline
     \end{tabular}
    \caption{Stellar parameters correspond to point B, point D, point O, point C, point B' and point E are shown above. Here the lines are star age, the mass of the star ($ M$), effective temperature ($ T_{\rm eff}$), surface luminosity ($ L$), surface radii ($ R $), surface gravity ($ g $), and mass loss via stellar winds ($ \dot{M}$) respectively.}
    \label{T1}
\end{table*}

\subsection{Giant Eruption}
\label{3.3}
The second simulation includes an induced massive mass loss event that emulates a GE.
Starting from a point in the post-MS (point B in Figure \ref{hr}), we introduced a mass loss rate of $  ~ 0.15~ {M_{\odot}}~{\rm yr^{-1}} $ for 20 years.
The starting point of the mass loss is chosen as the point where $T \simeq 19\,400 $ K, which corresponds to the place on the LBV instability strip for $L\simeq 10^6 {L_\odot}$  \citep{2017RSPTA.37560268S}. The choice to identify point B for the start of mass loss is backed by the observed placement of LBVs like $ \eta$- Carina  \citep[e.g.,][]{1999PASP..111.1124H, 2006AJ....132.2717M, 2007ApJ...666.1116S,2008MNRAS.390.1751K, 2010MNRAS.405.1924K, 2010ApJ...717L..22M, 2012Natur.486E...1D, 2012ApJ...751...73M, 2012ASSL..384....1H, 2014A&A...564A..14M} and P Cygni \citep[e.g.,][]{1988IrAJ...18..163D,1994PASP..106.1025H, 1997A&A...326.1117N,1999SSRv...90..493I, 1999ASPC..192...32D,2001ASPC..233..133N, 2010AJ....139.2269B, 2010MNRAS.405.1924K, 2012ASSL..384...43D,2018NewA...65...29M, 2020MNRAS.494..218R} in this region on the HR diagram, where they undergo the eruption phase and sheds a significant mass from their envelope.

The evolutionary track from point B to C by an orange line in Figure \ref{hr} represents the simulated GE phase.
The star evolves towards the hotter side of the HR diagram (see Figure \ref{hr}). This suggests mass loss suppresses inflation by reducing the temperature gradients closer to the surface. It removes the outer envelope to reveal the deeper, and hotter layers, which causes the star to evolve to the hotter temperature. This also affects the radius ($R$) evolution of the star. Thus, the mass loss rate has a significant impact on how the temperature of the stellar model evolves for the higher masses \citep{2021MNRAS.506.4473S, 2022MNRAS.514.3736S}. The \textsc{mlt++} option is turned off, and the only variable is the mass-loss scaling, utilized during the evolution, with all other inputs remaining constant in this phase. The trajectory of the star during the GE phase has two distinct parts.

The brightness increases from point B to O before abruptly decreasing from point O to C. As the star evolves, it undergoes a radial expansion for $\simeq 3.28$ days (from point B to O) and 
begins to contract over time (from point O to C).
Stellar parameters corresponding to points O and C (final stage of GE) are given in Table \ref{T1}. 
During the GE (from point B to C), as mass is being removed from the surface and new material is exposed as the outer layer.
This transition is accompanied by a slight increase in the temperature and a sharp drop in the luminosity.
The luminosity behaviour follows the value of $ \Gamma $.
Since the value of $ \Gamma $ is high and the transition satisfies the condition {$\eta $ = $ f \tau$} (where $\eta $ is a wind efficiency parameter and $ \tau $ is the sonic optical depth) the Eddington parameter initially exceeds unity and then decreases over time as the luminosity drops (see Figure \ref{Edd} panel (a) and section \ref{3.4} below).
The same behaviour was obtained by \citet[][see their figure 3]{2022MNRAS.514.3736S}.

The quiescence model, on the other hand, stays close to the Eddington limit (Figure \ref{Edd} (a)). \citet{2012ApJ...751L..34V} introduced an empirical correction factor $f$ for the transition between O and WR spectral type stars. They used a simple $ \rm \beta $-law model and found $f \approx 0.6 $ in their model. In our work, we satisfy both the Eddington parameter and the correction factor requirements. Here we are not considering any dependence on surface abundance and temperature, so the Eddington parameter depends on the $ L/M $ ratio only \citep{2000A&A...362..295V, 2001A&A...369..574V}. Based on this condition, as the star evolves from point B, its $ L/M$ ratio should be higher. This causes a spike in luminosity from point B to O and the star expands initially for $\simeq 3.28$ days. Further, as the star evolves with time, there is a sharp drop in the luminosity, as we can see in Figure \ref{hr}, from point O to C. It is caused by the loss of mass during the GE and we find out that the effective temperature remains almost constant throughout the evolution \citep{2022MNRAS.514.3736S}. Thus the temperature of the massive star decreases due to inflation and increases by mass loss.

 \begin{figure}
\centering\begin{tabular}{c}
\includegraphics[scale=0.52]{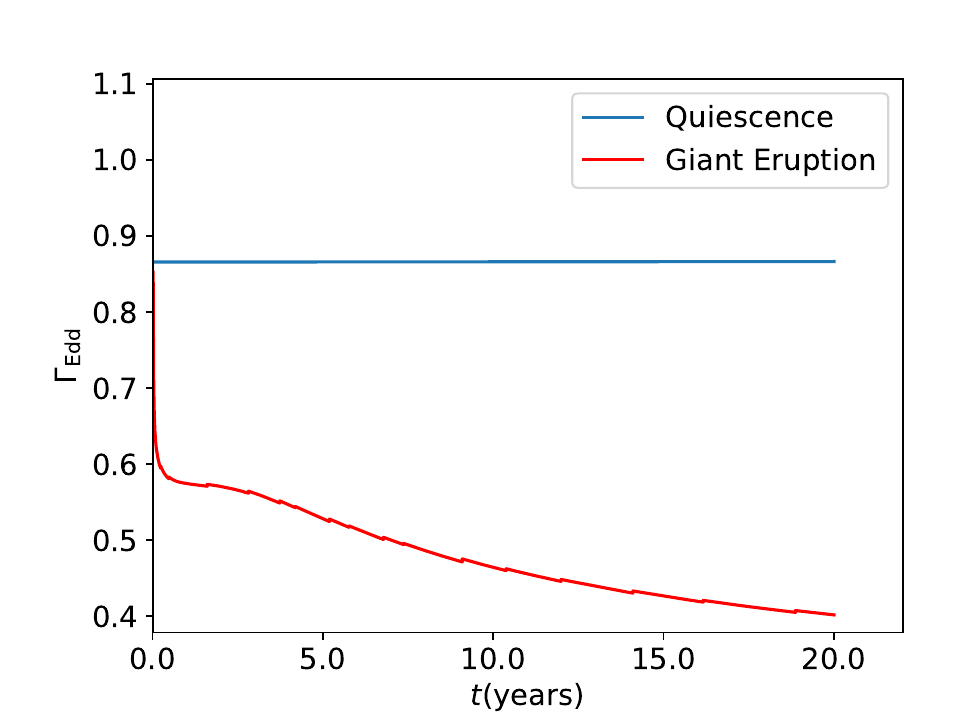}\\
(a)\\
\includegraphics[scale=0.52]{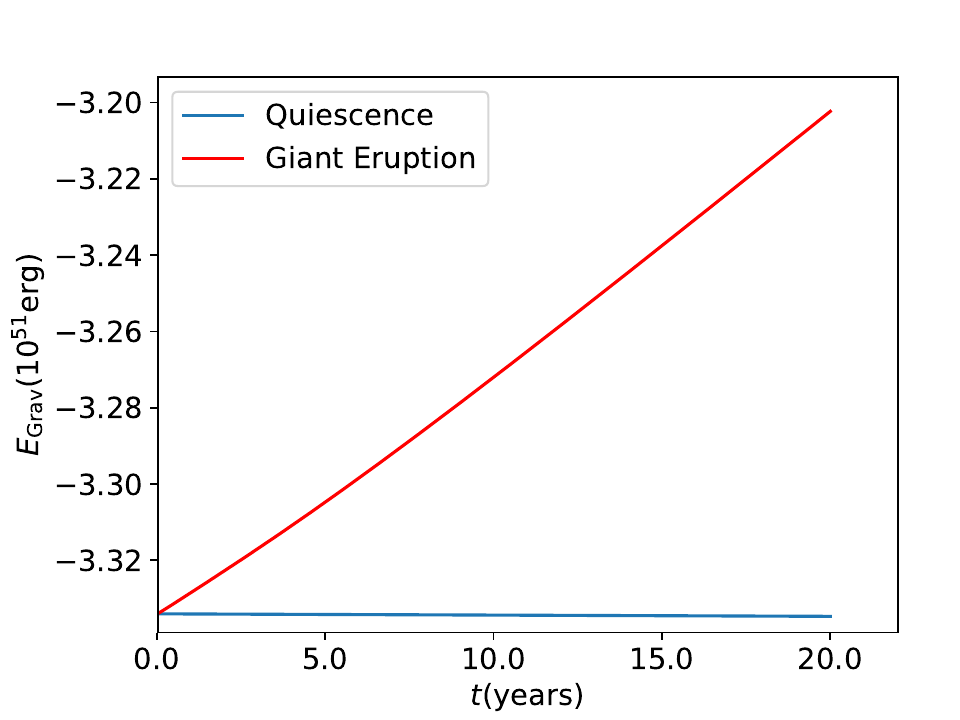}\\
(b)\\
\includegraphics[scale=0.52]{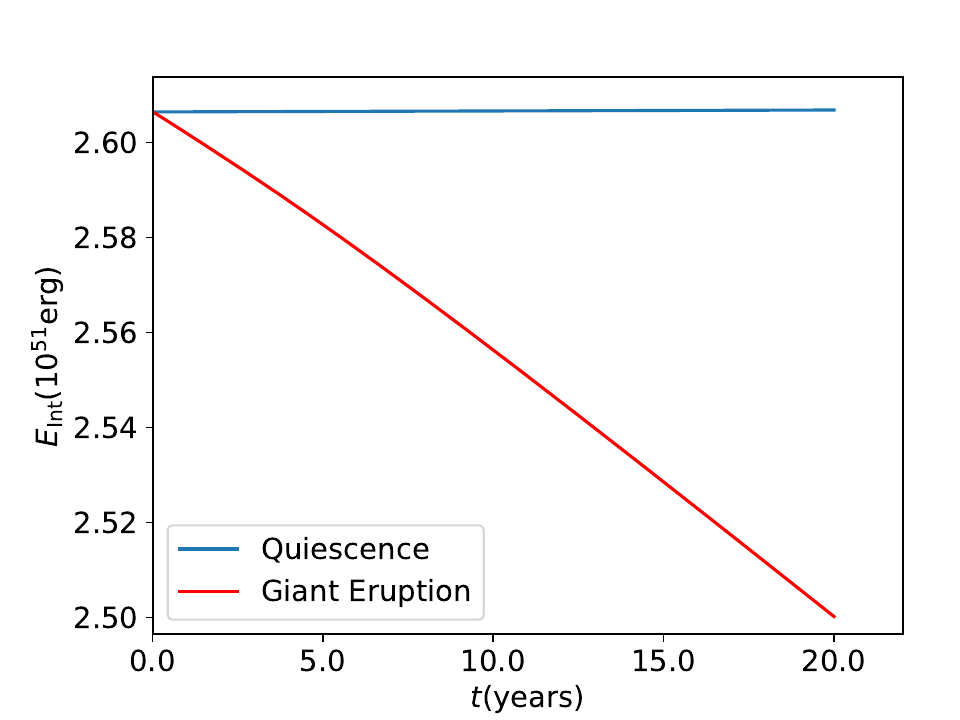}\\
(c)

\end{tabular}
\caption{
The variation with time starting from the GE of the Eddington factor $\Gamma $= $ L/{ L_{ \rm Edd}}$ (panel a), gravitational energy (panel b) and internal energy  (panel c). Here t=0 refers to point B where the star age is 2.918795 Myr, and t=20 refers to point C where the star age is 2.918815 Myr respectively as mentioned in Table \ref{T1}. During the simulated GE the  star loses mass at a rate of $0.15~ M_{\odot}~\rm yr^{-1} $. The quiescence case with no eruption is also shown, presenting constant values.}
\label{Edd}
\end{figure}

\subsection{The stellar surface during the GE}
\label{3.4}

Another way to think about the evolutionary tracks of $ 70~M_{\odot}$ during the quiescence and GE phase (Figure \ref{hr}) in terms of the change in luminosity is described below: 

By writing the equation of energy conservation we can express the specific gravitational energy $\epsilon_{\rm grav} $ in terms of the luminosity $ L_{\rm actual}$ and $  L_{\rm nuc}$
\begin{equation}
     \epsilon_{\rm grav} = \frac{ d L_{\rm actual}}{dm} - \frac{dL_{\rm nuc}}{ dm}  .
\end{equation}

On thermal timescale, a star can either shrink or expand as a result of variations in (i) the core temperature, core density, or fuel supply of a nuclear burning zone (which affects $ L_{\rm nuc}$). (ii) The hydrodynamic structure of the star (affects $ L_{\rm actual} $)
(iii) The efficiency of energy transport, such as how opacity changes or if convection is present, can affect $ L_{\rm actual}$.
The surface values of a star change if there is a mismatch between $ L_{\rm nuc} $ and $L_{\rm actual} $ due to a modification to the internal abundance profile, the hydrostatic structure, or the energy transit during the evolution.
As the star approaches thermal equilibrium, a new $L$ and $T_{\rm eff}$ form \citep{2022MNRAS.512.4116F}.

In the quiescence model, the star expands normally (from point B to point D in Figure \ref{hr}), and the expansion halts when the energy generated in the hydrogen shell is equal to the energy that the envelope can convey i.e., $ L_{\rm nuc}$ =  $ L_{\rm actual}$.
At point B the star is still at the stage where hydrogen is burned into helium increasing the mean molecular weight, and as a result, it is expanding.
Also, if the mass of the envelope decreases, it causes a decrease in radius, and high $T_{\rm eff} $ \citep{2020MNRAS.495.4659F, 2022MNRAS.512.4116F}.

In the GE model, we introduced an enhanced mass loss rate at point B, and thus the mass is removed from the outer layers of the star. In this instance, the star experiences two effects. During the initial $\simeq 3.28$ days of mass ejection, the effect of high mean molecular weight dominates, and as a response, the star expands initially from point B to O (see Figure \ref{hr}). The star begins to contract (from point O to point C) when a sizable quantity of hydrogen is ejected, allowing helium to increase in abundance and to reduce the opacity,  which causes the smaller radii, higher $ T_{\rm eff}$, and lower surface luminosity.

 \begin{figure*}
  \centering
  \begin{tabular}{c @{\qquad} c }
    \includegraphics[width=.52\linewidth]{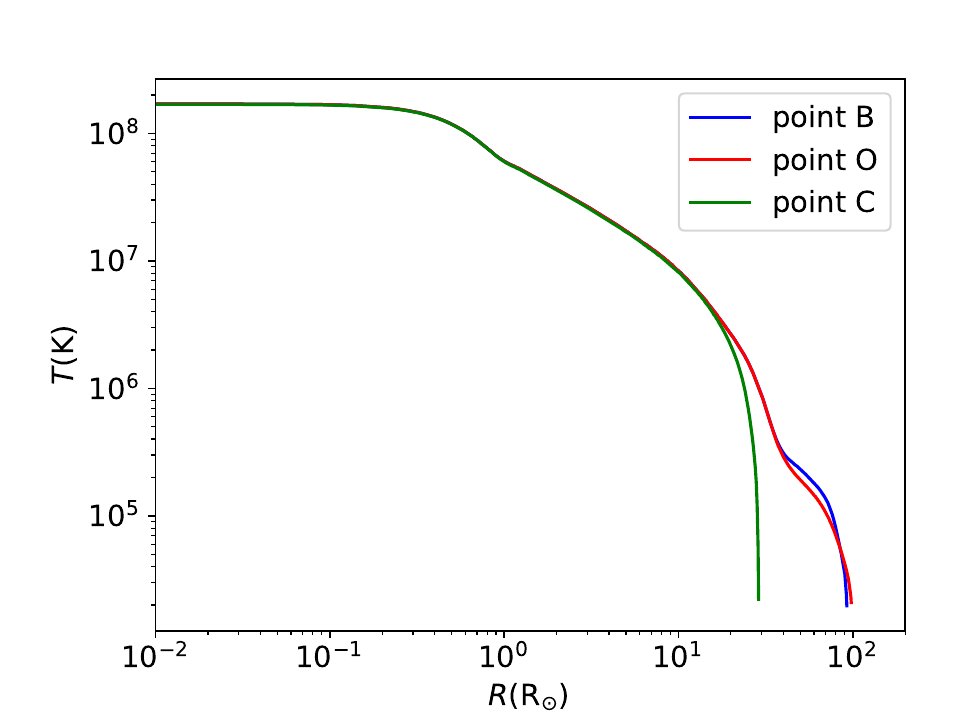}
    &
     \hspace{-1.2cm} 
    \includegraphics[width=.52\linewidth]{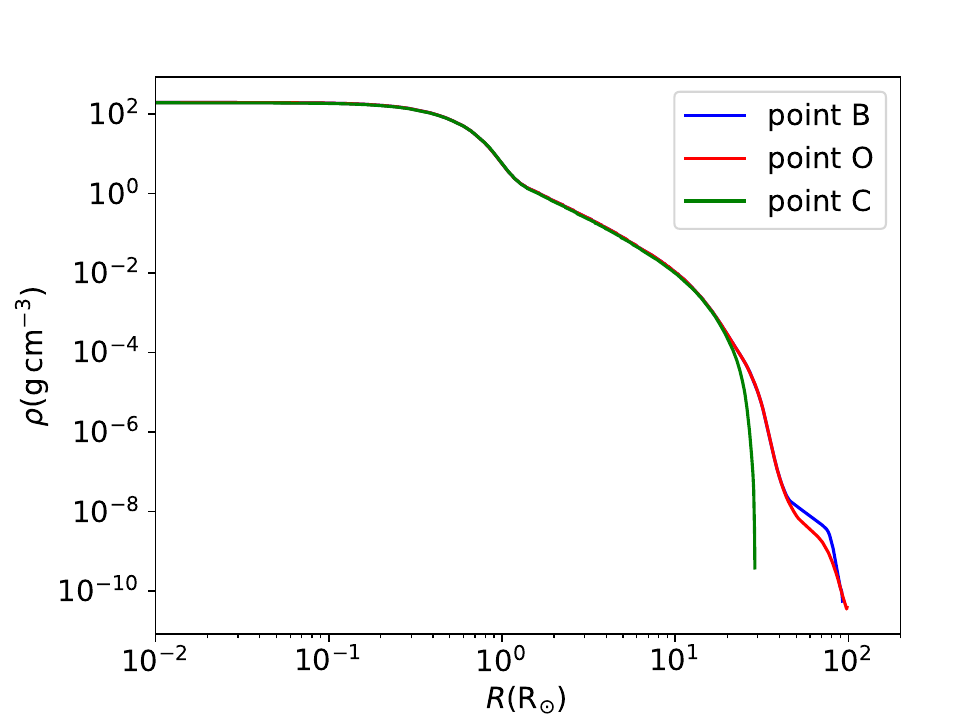}
    \\
    \small (a) & (b)
  \end{tabular}

\caption{
Temperature (panel a) and density (panel b) profiles at three different time points (B, O, and C; Figure \ref{hr}) during the simulated GE.
The star initially shows expansion and a mild increase in temperature and density from point B to point O, followed by contraction till point C.
}

  \label{TRho}
\end{figure*}
    \begin{figure*}
  \centering
  \begin{tabular}{c @{\qquad} c }
    \includegraphics[width=.52\linewidth]{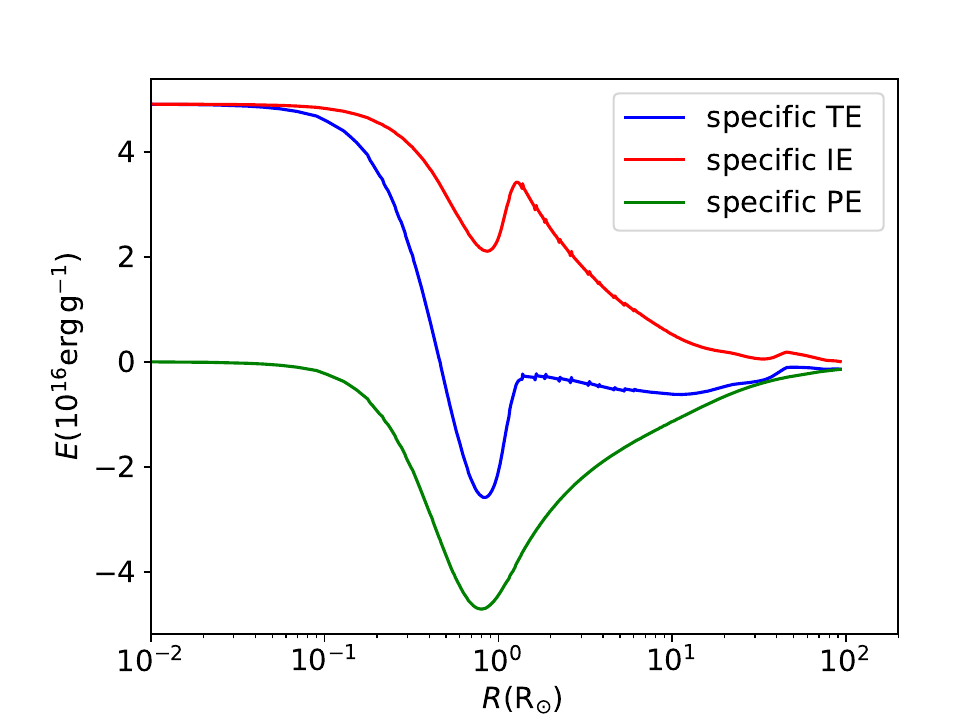}
    &
     \hspace{-1.2cm} 
    \includegraphics[width=.52\linewidth]{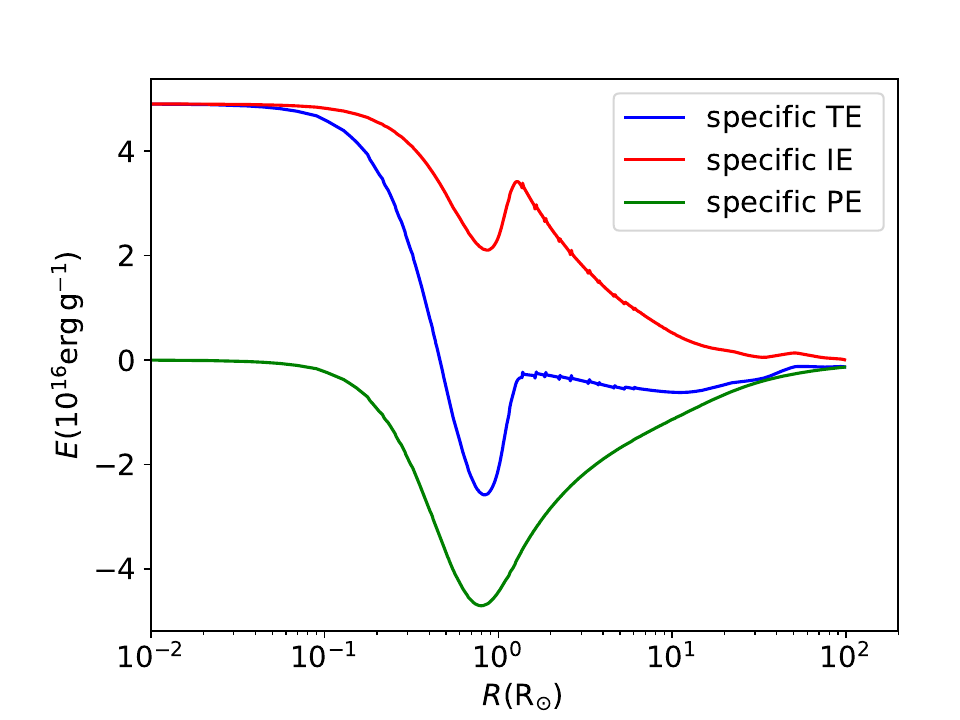}
    \\
    \small (a) & (b)
  \end{tabular}
  \centering
   \begin{tabular}{c @{\qquad} c }
    \includegraphics[width=.52\linewidth]{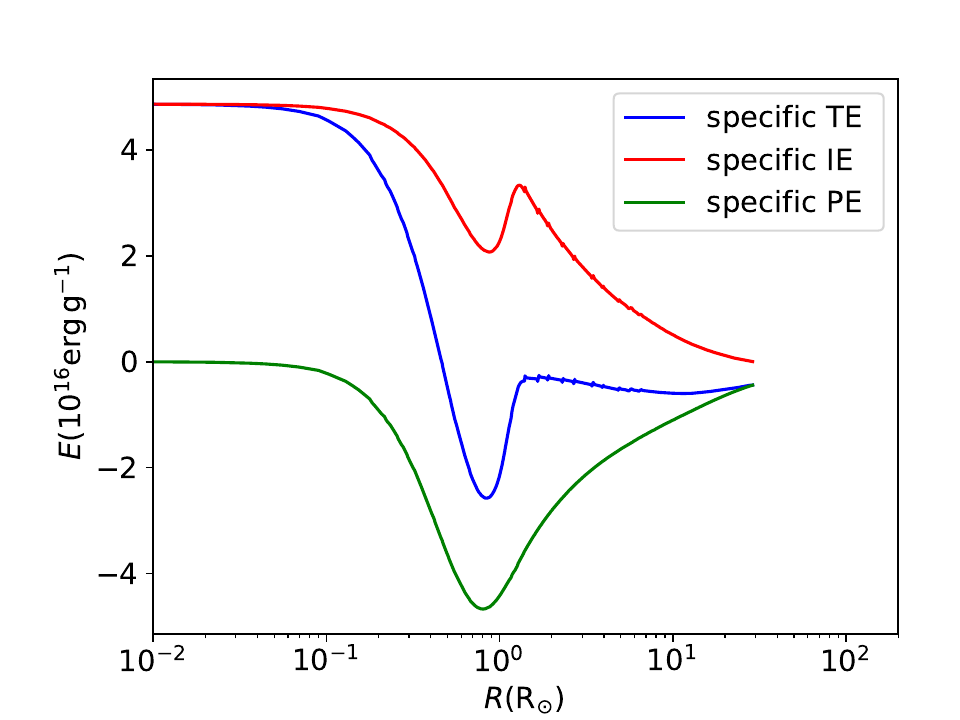}
    &
     \hspace{-1.2cm} 
    \includegraphics[width=.52\linewidth]{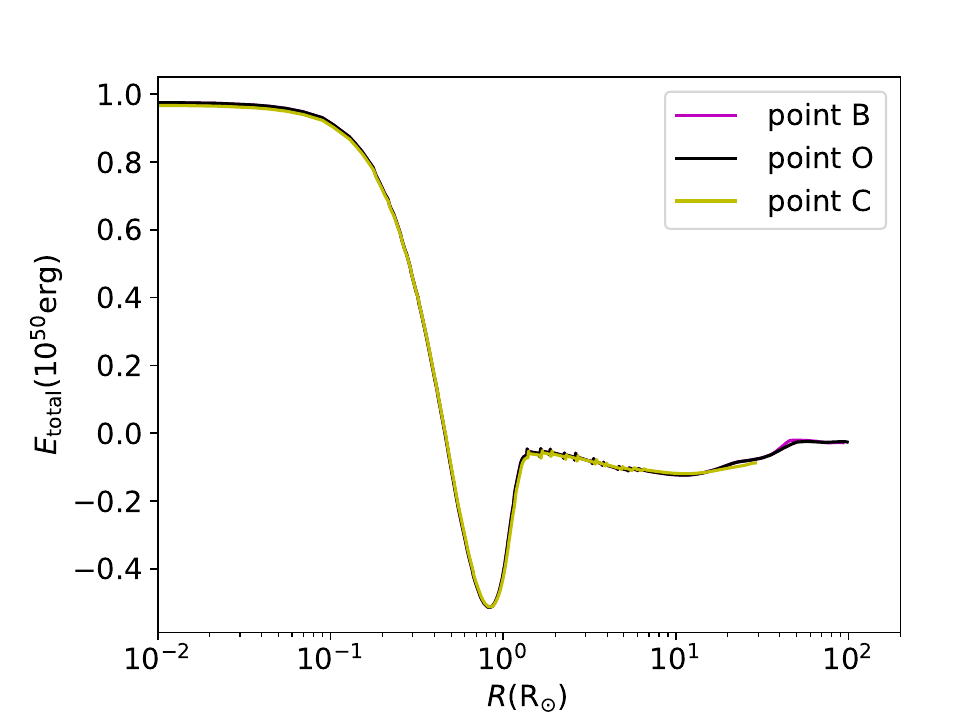}
    \\
    \small (c) & (d)
  \end{tabular}
  
\caption{ The internal profile of specific total energy, internal energy, and gravitational energy from the star's surface to its core at three distinct locations (B, O, and C, respectively) as the function of stellar radii is shown above in panels (a), (b), and (c). Panel (d) shows the total energy of the star as a function of the stellar radii. The star initially expands after introducing mass loss, and its internal energy decreases during the 20 years. But both factors are not sufficient enough to lower the temperature of the core significantly (see Figure \ref{3} (a)) when the star progresses from point B to point C. As we can see in panel (d), it makes a negligible difference in the energy production inside the core.}
 
  \label{eall}
\end{figure*}
 \begin{figure*}
  \centering
  \begin{tabular}{c @{\qquad} c }
    \includegraphics[trim= 0.0cm 0.0cm 0.0cm 1.2cm,clip=true,width=0.52\linewidth]{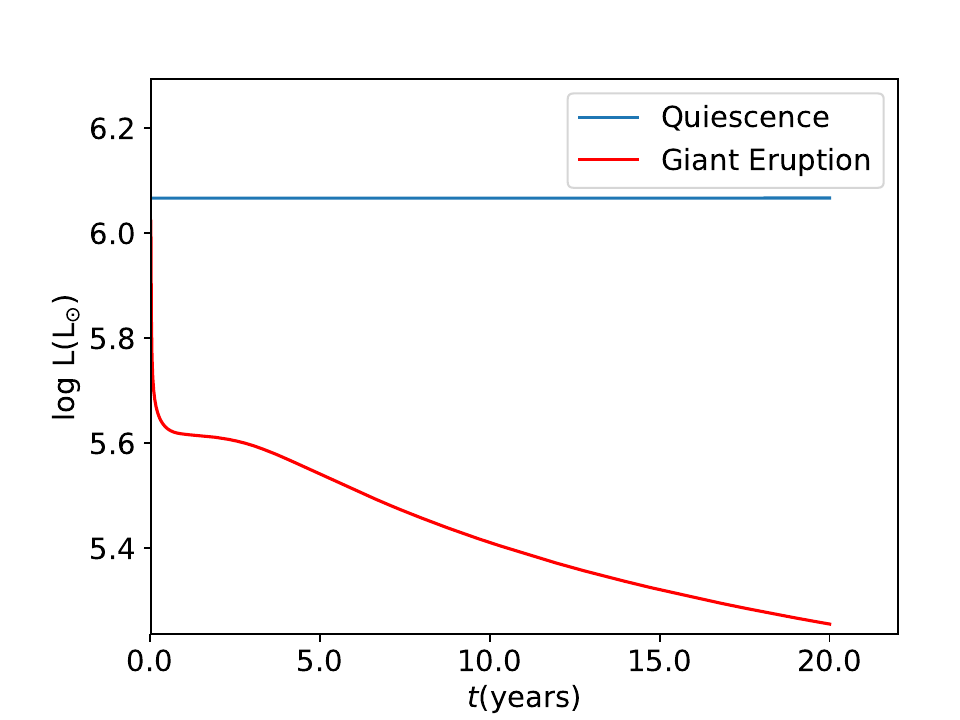}
    &
     \hspace{-1.2cm} 
    \includegraphics[trim= 0.0cm 0.0cm 0.0cm 1.2cm,clip=true,width=0.52\linewidth]{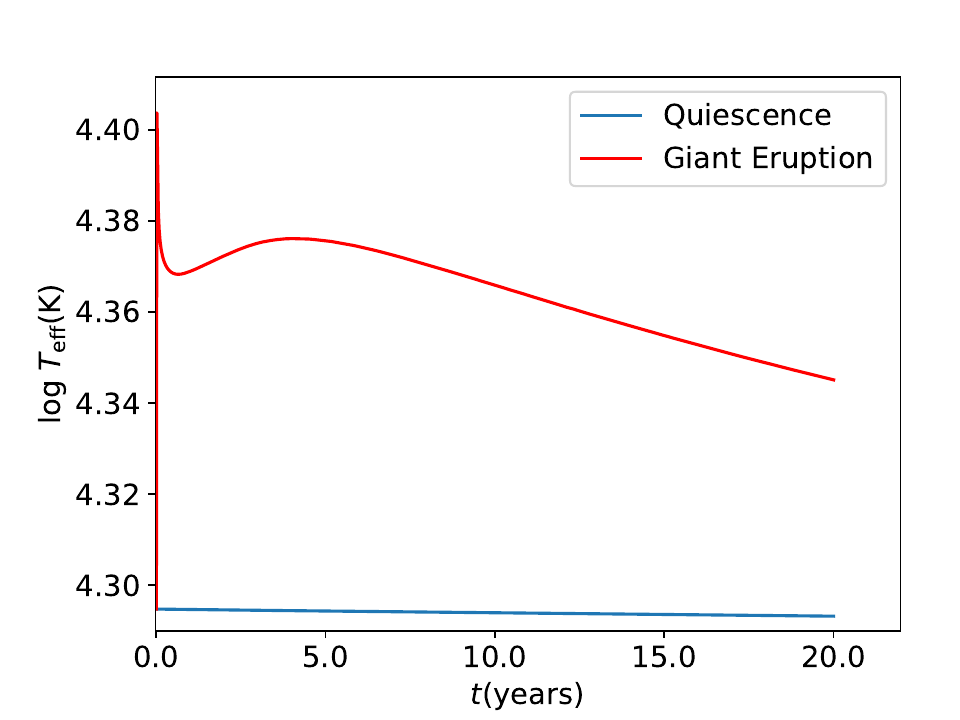}
    \\
    \small (a) & (b)
  \end{tabular}

  \caption{Panels (a) and (b) show the variation in luminosity and temperature. In quiescence, relatively few changes are seen in luminosity. Yet, in the GE, it first increases based on the state of transitional mass loss and decreases later on as the release of gravitational energy. During the 20 years of mass loss, the temperature has remained almost constant (see Figure \ref{hr}.)}
 
  \label{all}
\end{figure*}

Figure \ref{Edd} panels (b) and (c) illustrate the gravitational and internal energy distribution throughout the 20 years of mass loss. Since mass is depleted from the outermost surface layer to the innermost, gravitational energy increases, as a result, internal energy declines to conserve the total energy of the system. In panel (b),  the gravitational energy ($ E_{\rm Grav}$) increases and becomes more significant in $ \rm 20~ years $ of mass loss. At beginning of the eruption t = 0 (point B), gravitational energy is $ \rm -3.33 $ $\rm  \times 10^{51}$ erg. At t = 20 years (point C) of mass loss, the gravitational energy is $ \rm -3.20 $ $\rm  \times 10^{51}$ erg.  Similarly. in panel (c) at star age, t = 0 (point B), the internal energy is $ \rm 2.60 $ $\rm  \times 10^{51}$ erg after t = 20 years (point C) of the mass loss, internal energy ($ E_{\rm Int}$) is $ \rm 2.50 $ $\rm  \times 10^{51}$ erg. The gravitational energy that is lost throughout the 20 years of evolution (GE) from point B to point C is $\Delta E_{\rm Grav} $ = $\rm 1.30$ $\rm  \times 10^{50}$ erg. This lost energy is almost equivalent to that which was lost during the GE of $\eta $ Carinae in the 19$^{\rm th}$ century \citep{2008MNRAS.390.1751K, 2012NewA...17..616S,2013MNRAS.429.2366S, 2019MNRAS.486..926K}. Similarly, the internal energy falls by the value of $ \Delta E_{\rm Int} $ = $\rm 1.0$ $\rm  \times 10^{50}$ erg.

In Figure \ref{TRho} (a) and (b), the temperature and density profiles are depicted, revealing that during the mass loss process, the star's interior remains relatively constant. There are negligible differences in temperature and density at points B, O, and C in the interior. The stellar radii vary at these points in the outer region due to the expulsion of outer layers. Specifically, the radius is larger at point O, as the star initially expands in response to the mass loss and becomes smaller at point C. The temperature of outer layers, impacted by a loss of internal energy and an early expansion, has a negligible impact on the energy generation rate in the star's interior. Both the temperature, and density rises from the surface to the interior as we approach inward, and it begins to stabilize between 0.1--1 $R_{\odot}$.

The radial profile of gravitational, internal, and total energy for a specific cell from the outer region to the interior of the star is shown in Figure \ref{eall} (a), (b), and (c) for points B, O, and C, respectively.
It can be seen that when the star contracts, the specific total energy (TE), specific internal energy (IE), and specific potential energy (PE) all decrease with radius down to a minimal value, and then increase towards the surface. Deep in the core they are almost constant. The total energy allocated to the points B, O, and C is shown in Figure \ref{eall} (d), and shows that the impact on energy production within the core is minimal due to the mass loss mechanism. The variation in the luminosity ($L$) temperature ($T_{\rm eff}$) during the GE, is depicted in Figure \ref{all}. During the GE phase, the luminosity decreases by a factor of $\simeq 1.4$,
and the temperature value increases by a factor of $\simeq 1.01$ (see Table \ref{T1}).

\begin{figure*}
  \centering
  \begin{tabular}{c @{\qquad} c }
    \includegraphics[trim= 0.0cm 0.0cm 0.0cm 0.7cm,clip=true,width=0.52\linewidth]{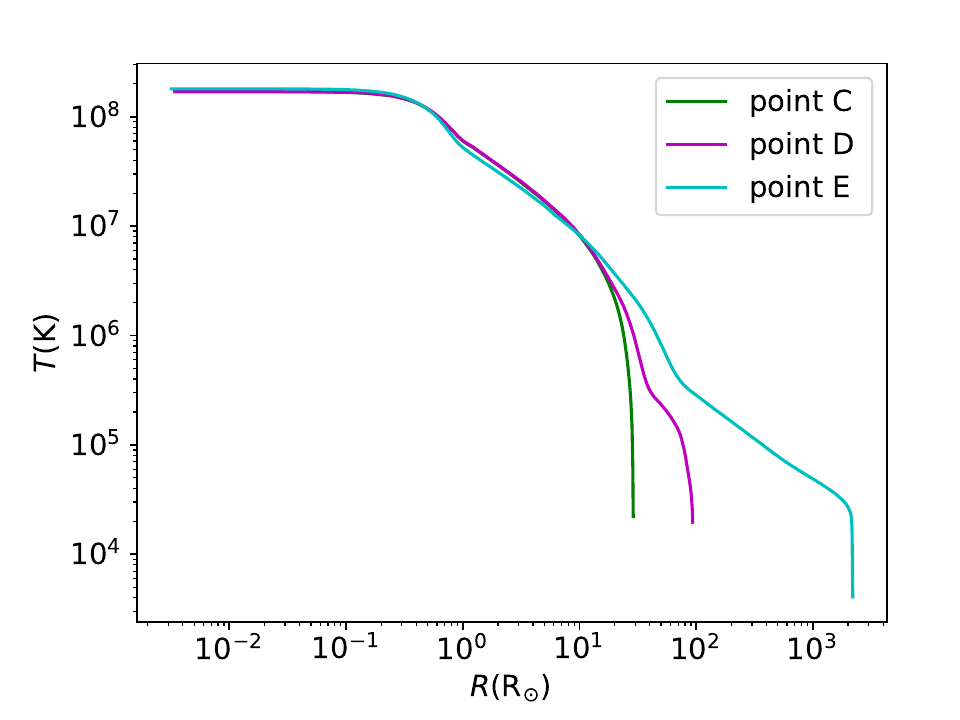} %trim=l b r t
    &
     \hspace{-1.2cm} 
    \includegraphics[trim= 0.0cm 0.0cm 0.0cm 0.7cm,clip=true,width=0.52\linewidth]{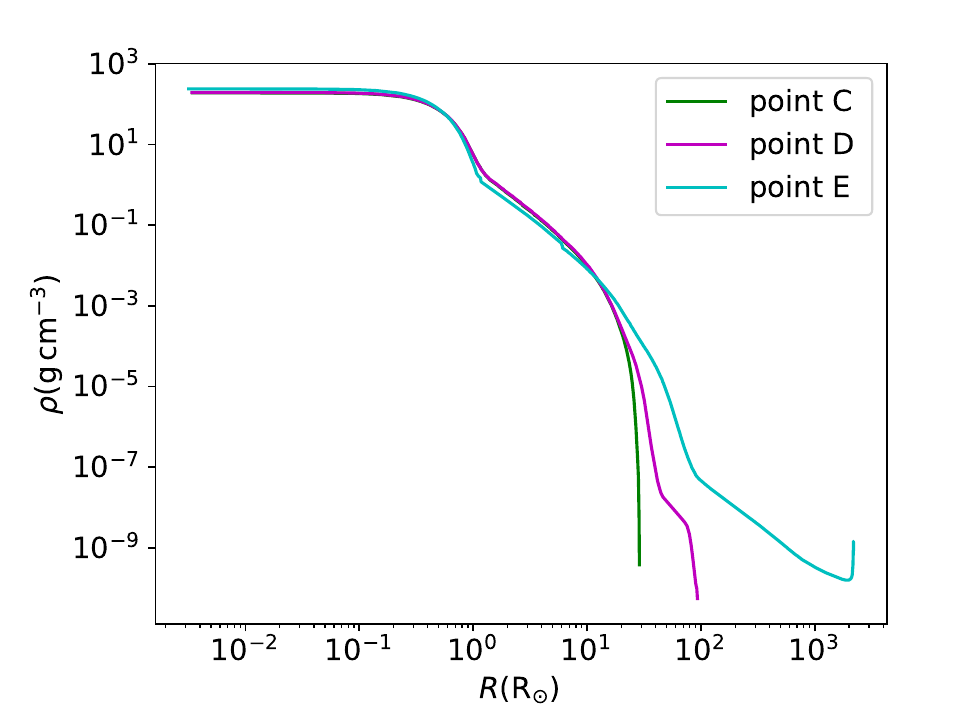}
    \\
    \small (a) & (b)
  \end{tabular}

\caption{The temperature and density profile from the star's surface to its core at three distinct locations C, D, and E, respectively, as 
the function of stellar radii is shown in panels (a) and (b).Here the last profiles of the GE are represented by point C, the quiescence by point D, and the star's recovery by point E. In the quiescence, inflation occurs and as a result point, D has a larger radius as compared to points C, and E. Temperature profiles demonstrate that after point C, the star expands and as a result, point E has larger radii than point C. As we can see in the density profiles, the density at point E is slightly higher than at point C.
During the recovery of a star, the central regions of the star become denser while the outer regions become less dense. This is because the core of the star has depleted its nuclear fuel and is no longer generating enough energy to counteract the gravitational forces pulling the star inward.}
 
  \label{TRcde}
\end{figure*}

\subsection{Recovery}
The recovery period of a star following a GE is typically prolonged, often extending over many years or even decades, as found in earlier works \citep[e.g.,][]{2009NewA...14..539H, 2012ASSL..384...43D, 2016ApJ...817...66K}.
We followed the star evolution form the GE for a total duration of 3359 years. It is represented from point C to E by the red dashed line in Figure \ref{hr}.
We find that it took the star $\simeq 87$ years to arrive at a state (point B' in Figure \ref{hr} i.e., the end phase of the recovery) similar to its state before the eruption  (point B). This can be regarded as the time it taked the star to recover from the eruption.
From point B' to point E we continued to follow its evolution (3272 years).

During the evolution of the star, after the eruption (i.e., track from point C to B' and B' to E), the star's outer layers, which were exposed after the layers above them had been expelled during the eruption, changed in size, temperature, and luminosity. Although the mass loss from the star's outer layers affects its structure, the internal region during the GE and recovery remains stable, as shown in Figure \ref{TRcde} (a) and (b). At points C, D, and E in the stellar interior, the temperature and density remain relatively constant, yet in the outer regions, these parameters undergo significant changes.
As shown in Figure \ref{TRcde}, the final profiles of the GE are denoted by point C, the quiescent state by point D, and the star's recovery by point E. During quiescence, inflation occurs, resulting in point D having a larger radius compared to points C and E. The temperature profiles in Figure \ref{TRcde} panel (a) indicate that after point C, the star expands, leading to point E having a larger radius than point C. As observed in the density profiles in Figure \ref{TRcde} panel (b), the density at point E is slightly higher than at point C.

During the recovery phase and further evolution (track from B' to E), there are two jumps or loops near $\log T_{\rm eff}= 4.35$ K and $\log T_{\rm eff}= 3.8$ K. They are a result of a large time step that was a necessity for this long duration of evolution, and do not represent real variations. They may slightly affect the position of points B' and E.
Stellar parameters corresponding to points B' and E (final stage of recovery) are given in Table \ref{T1}. Figure \ref{hr} shows that during the recovery phase, the star undergoes evolution toward the cooler side of the HR diagram. During this phase, mass redistribution occurs and the star evolves towards a cooler side of the HR diagram, similar to the quiescence phase, where it gradually regains equilibrium.

\section{Conclusions and Summary}
\label{4}

We used \textsc{mesa} to track the evolution of a $70~M_{\odot} $ star undergoing a giant eruption by removing $ 3~ M_{\odot} $ from the outer layers of the star over 20 years at a constant rate. We examined the impact of mass loss on the stellar parameters in the aftermath of the eruption and obtained a physical transition where the evolutionary track switches from the cool side to the hot side of the HR diagram \citep{2012ApJ...751L..34V, 2022MNRAS.514.3736S}.
During the simulated GE, the radius shrinks by a factor of 1.34 and the loss of gravitational potential energy is $\simeq 1.3 \times 10^{50}$ erg, which constitutes $\simeq 5$ \% of the total value. %10.94
It is usually expected that high mass loss rates will bring the star to higher temperatures, however, the temperature of the star remained almost unchanged from point O to C while the star experienced a sharp decline in luminosity (see Figure \ref{hr}). The reason is that the mass loss rate is directly correlated with the star's luminosity \citep{2001A&A...369..574V}, and a significant decrease in luminosity counteracts the effect of mass loss, resulting in an almost constant temperature.

It took $\simeq 87$ years for the star to reach back from point C after the eruption to very close to point B (point B' in Figure \ref{hr}),
though a different starting point B would have resulted in a different point of intersection.
A higher mass loss rate during the GE would probably shorten the time it takes for the star to reach B', namely a recurrent GE would occur on a shorter timescale.
The steeper slope at B' suggests a more rapid evolutionary transition.
We propose that at this point, the star might undergo another GE.
If the system is binary, and the eruptions are induced by the gravitational force of the companion in an eccentric orbit \citep[e.g.,][]{2010MNRAS.405.1924K}, the time of the second GE may depend on the orbital position of the secondary and start at a later time, or be avoided.
Simulating multiple GEs is beyond the scope of the present paper.
We thus do not simulate another GE, but rather let the star continue its evolution.
The recovering star later continues from point B' to point E on a trajectory that lasts $\simeq 3272$ years. Its luminosity increases slowly and it expands and cools. As a result, after passing the recovery phase (point B') it sets to the more luminous position (point E) on the HRD.

The star we simulated crossed the Humphreys-Davidson (HD) limit for both the quiescence and the GE simulations.
For the quiescence simulation, this is of course expected as there is no mass loss to prevent the star from crossing it.
For the GE simulation, as explained, we did not model the evolution of a star that might have stronger or multiple GEs that would prevent it from crossing the HD limit.
The numerical experiment we are doing is only an assumed isolated GE that does not necessarily represent the entire history of events the star experiences during its evolution. We note that to avoid the crossing of the HD limit for a $ 70~M_{\odot}$ star, \textsc{MESA} introduced a modification version of the mixing-length theory (\textsc{mlt++}) that allows massive stars to evolve towards higher temperatures before reaching the HD limit \citep{2013ApJS..208....4P}.

The simulated star shows different behavior to GEs of observed GEs, as usually GEs are events of an increase in luminosity, while the star we simulated shows an initial increase but then a sharp decrease in luminosity of more than one order of magnitude.
The reason is that the energy of the GE does not come from the LBV, but rather from accretion onto the companion star \citep[e.g.,][]{2001MNRAS.325..584S, 2010ApJ...723..602K, 2016RAA....16..117S}.
The gravitational energy released by the accretion onto the companion powers the GE and results in the high luminosity brightening, which appears to be crossing the Eddington value.
Another different behavior is that the stellar temperature we obtain remains hot during the eruption, unlike observed GEs that show cooler temperature \citep{1987ApJ...317..760D}. However, we simulate only the star and not the observed properties through the ejecta, neither we model a pseudo-photosphere, so they are not expected to be similar.

Massive star evolution is strongly tied to mass loss, emphasizing the importance of accurate modeling and understanding of this process. The GE we simulated here is an example of such a process that can occur for a wide range of massive stars and magnitudes. Yet, GEs are rare events, suggesting that a set of conditions is required for them to occur. In a future work, we will employ varying mass loss rates over different time intervals, and for a range of massive stars, covering the parameter space to quantitatively understand the impact of GEs on the evolution of massive stars.

\begin{acknowledgments}
We thank the anonymous referee whose comments and suggestions have helped to improve this work.
We thank Dina Prialnik and Noam Soker for very helpful comments.
We acknowledge support from the R\&D authority at Ariel University.
We acknowledge the Ariel HPC Center at Ariel University for providing computing resources that have contributed to the research results reported in this paper.
The MESA inlists and input files to reproduce our simulations and associated data products are available on Zenodo (DOI: 10.5281/zenodo.13171005).
All figures are made with the free Python module Matplotlib (\url{https://matplotlib.org}).
\end{acknowledgments}

\newpage

\section*{APPENDIX}
\label{app}
The parameters used in our simulation are presented in this section:
\newline \texttt{max}$\texttt{\_timestep}$ = 1000
\newline \texttt{mesh}$\texttt{\_delta}$$\texttt{\_coeff}$ = 1.5
\newline \texttt{time}$\texttt{\_delta}$$\texttt{\_coeff}$ = 1
\newline \texttt{varcontrol}$\texttt{\_target}$ = 1d-3

From the \textsc{mesa} control defaults, we added a few extra parameters to our inlist file to account for the 20 years of mass loss evolution. These conditions are:
\newline \texttt{mass}$\texttt{\_change}$ = -0.15 
\newline \texttt{remnant}$\texttt{\_mass}$$\texttt{\_min}$$\texttt{\_limit} $  = 67.00
\newline \texttt{Ejecta}$\texttt{\_mass}$$\texttt{\_max}$$\texttt{\_limit}$ = 3

The boundary conditions for the atmosphere are:
\newline \texttt{atm}$\texttt{\_option}$ = $\texttt{T}$ $\texttt{\_tau}$
\newline \texttt{atm}$\texttt{\_T}$$\texttt{\_tau}$$\texttt{\_relation}$ = $\texttt{Eddington}$ 
\newline \texttt{atm}$\texttt{\_T}$ $\texttt{\_tau}$$\texttt{\_opacity}$ = $\texttt{varying} $

\bibliography{GE_Massive}

\begin{thebibliography}{}
\makeatletter
\relax
\def\mn@urlcharsother{\let\do\@makeother \do\$\do\&\do\#\do\^\do\_\do\%\do\~}
\def\mn@doi{\begingroup\mn@urlcharsother \@ifnextchar [ {\mn@doi@} {\mn@doi@[]}}
\def\mn@doi@[#1]#2{\def\@tempa{#1}\ifx\@tempa\@empty \href {http://dx.doi.org/#2} {doi:#2}\else \href {http://dx.doi.org/#2} {#1}\fi \endgroup}
\def\mn@eprint#1#2{\mn@eprint@#1:#2::\@nil}
\def\mn@eprint@arXiv#1{\href {http://arxiv.org/abs/#1} {{\tt arXiv:#1}}}
\def\mn@eprint@dblp#1{\href {http://dblp.uni-trier.de/rec/bibtex/#1.xml} {dblp:#1}}
\def\mn@eprint@#1:#2:#3:#4\@nil{\def\@tempa {#1}\def\@tempb {#2}\def\@tempc {#3}\ifx \@tempc \@empty \let \@tempc \@tempb \let \@tempb \@tempa \fi \ifx \@tempb \@empty \def\@tempb {arXiv}\fi \@ifundefined {mn@eprint@\@tempb}{\@tempb:\@tempc}{\expandafter \expandafter \csname mn@eprint@\@tempb\endcsname \expandafter{\@tempc}}}

\bibitem[\protect\citeauthoryear{{Abbott}}{{Abbott}}{1982}]{1982ApJ...259..282A}
{Abbott} D.~C.,  1982, \mn@doi [\apj] {10.1086/160166}, \href {https://ui.adsabs.harvard.edu/abs/1982ApJ...259..282A} {259, 282}

\bibitem[\protect\citeauthoryear{{Aghakhanloo} et~al.,}{{Aghakhanloo} et~al.}{2023a}]{2023MNRAS.tmp.2585A}
{Aghakhanloo} M.,  et~al., 2023a, \mn@doi [\mnras] {10.1093/mnras/stad2702}, \href {https://ui.adsabs.harvard.edu/abs/2023MNRAS.tmp.2585A} {}

\bibitem[\protect\citeauthoryear{{Aghakhanloo} et~al.,}{{Aghakhanloo} et~al.}{2023b}]{2023MNRAS.521.1941A}
{Aghakhanloo} M.,  et~al., 2023b, \mn@doi [\mnras] {10.1093/mnras/stad630}, \href {https://ui.adsabs.harvard.edu/abs/2023MNRAS.521.1941A} {521, 1941}

\bibitem[\protect\citeauthoryear{{Agrawal}, {Stevenson}, {Sz{\'e}csi}  \& {Hurley}}{{Agrawal} et~al.}{2022}]{2022A&A...668A..90A}
{Agrawal} P.,  {Stevenson} S.,  {Sz{\'e}csi} D.,   {Hurley} J.,  2022, \mn@doi [\aap] {10.1051/0004-6361/202244044}, \href {https://ui.adsabs.harvard.edu/abs/2022A&A...668A..90A} {668, A90}

\bibitem[\protect\citeauthoryear{{Akashi} \& {Kashi}}{{Akashi} \& {Kashi}}{2020}]{2020MNRAS.494.3186A}
{Akashi} M.,  {Kashi} A.,  2020, \mn@doi [\mnras] {10.1093/mnras/staa1014}, \href {https://ui.adsabs.harvard.edu/abs/2020MNRAS.494.3186A} {494, 3186}

\bibitem[\protect\citeauthoryear{{Balan}, {Tycner}, {Zavala}, {Benson}, {Hutter}  \& {Templeton}}{{Balan} et~al.}{2010}]{2010AJ....139.2269B}
{Balan} A.,  {Tycner} C.,  {Zavala} R.~T.,  {Benson} J.~A.,  {Hutter} D.~J.,   {Templeton} M.,  2010, \mn@doi [\aj] {10.1088/0004-6256/139/6/2269}, \href {https://ui.adsabs.harvard.edu/abs/2010AJ....139.2269B} {139, 2269}

\bibitem[\protect\citeauthoryear{{Bj{\"o}rklund}, {Sundqvist}, {Puls}  \& {Najarro}}{{Bj{\"o}rklund} et~al.}{2021}]{2021A&A...648A..36B}
{Bj{\"o}rklund} R.,  {Sundqvist} J.~O.,  {Puls} J.,   {Najarro} F.,  2021, \mn@doi [\aap] {10.1051/0004-6361/202038384}, \href {https://ui.adsabs.harvard.edu/abs/2021A&A...648A..36B} {648, A36}

\bibitem[\protect\citeauthoryear{{Chiosi} \& {Maeder}}{{Chiosi} \& {Maeder}}{1986}]{1986ARA&A..24..329C}
{Chiosi} C.,  {Maeder} A.,  1986, \mn@doi [\araa] {10.1146/annurev.aa.24.090186.001553}, \href {https://ui.adsabs.harvard.edu/abs/1986ARA&A..24..329C} {24, 329}

\bibitem[\protect\citeauthoryear{{Cur{\'e}} \& {Araya}}{{Cur{\'e}} \& {Araya}}{2023}]{2023Galax..11...68C}
{Cur{\'e}} M.,  {Araya} I.,  2023, \mn@doi [Galaxies] {10.3390/galaxies11030068}, \href {https://ui.adsabs.harvard.edu/abs/2023Galax..11...68C} {11, 68}

\bibitem[\protect\citeauthoryear{{Davidson}}{{Davidson}}{1987}]{1987ApJ...317..760D}
{Davidson} K.,  1987, \mn@doi [\apj] {10.1086/165324}, \href {https://ui.adsabs.harvard.edu/abs/1987ApJ...317..760D} {317, 760}

\bibitem[\protect\citeauthoryear{{Davidson}}{{Davidson}}{2012}]{2012ASSL..384...43D}
{Davidson} K.,  2012, in {Davidson} K.,  {Humphreys} R.~M.,  eds,  Astrophysics and Space Science Library Vol. 384, Eta Carinae and the Supernova Impostors. p.~43, \mn@doi{10.1007/978-1-4614-2275-4_3}

\bibitem[\protect\citeauthoryear{{Davidson}}{{Davidson}}{2016}]{2016JPhCS.728b2008D}
{Davidson} K.,  2016, in Journal of Physics Conference Series. p. 022008 (\mn@eprint {arXiv} {1602.03925}), \mn@doi{10.1088/1742-6596/728/2/022008}

\bibitem[\protect\citeauthoryear{{Davidson} \& {Humphreys}}{{Davidson} \& {Humphreys}}{1997}]{1997ARA&A..35....1D}
{Davidson} K.,  {Humphreys} R.~M.,  1997, \mn@doi [\araa] {10.1146/annurev.astro.35.1.1}, \href {https://ui.adsabs.harvard.edu/abs/1997ARA&A..35....1D} {35, 1}

\bibitem[\protect\citeauthoryear{{Davidson} \& {Humphreys}}{{Davidson} \& {Humphreys}}{2012a}]{2012ASSL..384.....D}
{Davidson} K.,  {Humphreys} R.,  eds, 2012a, {Eta Carinae and the Supernova Impostors}  Astrophysics and Space Science Library Vol. 384, \mn@doi{10.1007/978-1-4614-2275-4.
}

\bibitem[\protect\citeauthoryear{{Davidson} \& {Humphreys}}{{Davidson} \& {Humphreys}}{2012b}]{2012Natur.486E...1D}
{Davidson} K.,  {Humphreys} R.~M.,  2012b, \mn@doi [\nat] {10.1038/nature11166}, \href {https://ui.adsabs.harvard.edu/abs/2012Natur.486E...1D} {486, E1}

\bibitem[\protect\citeauthoryear{{Dessart}, {Livne}  \& {Waldman}}{{Dessart} et~al.}{2010}]{2010MNRAS.405.2113D}
{Dessart} L.,  {Livne} E.,   {Waldman} R.,  2010, \mn@doi [\mnras] {10.1111/j.1365-2966.2010.16626.x}, \href {https://ui.adsabs.harvard.edu/abs/2010MNRAS.405.2113D} {405, 2113}

\bibitem[\protect\citeauthoryear{{Drake} et~al.,}{{Drake} et~al.}{2012}]{2012ATel.4334....1D}
{Drake} A.~J.,  et~al., 2012, The Astronomer's Telegram, \href {https://ui.adsabs.harvard.edu/abs/2012ATel.4334....1D} {4334, 1}

\bibitem[\protect\citeauthoryear{{Farrell}, {Groh}, {Meynet}, {Eldridge}, {Ekstr{\"o}m}  \& {Georgy}}{{Farrell} et~al.}{2020}]{2020MNRAS.495.4659F}
{Farrell} E.~J.,  {Groh} J.~H.,  {Meynet} G.,  {Eldridge} J.~J.,  {Ekstr{\"o}m} S.,   {Georgy} C.,  2020, \mn@doi [\mnras] {10.1093/mnras/staa1360}, \href {https://ui.adsabs.harvard.edu/abs/2020MNRAS.495.4659F} {495, 4659}

\bibitem[\protect\citeauthoryear{{Farrell}, {Groh}, {Meynet}  \& {Eldridge}}{{Farrell} et~al.}{2022}]{2022MNRAS.512.4116F}
{Farrell} E.,  {Groh} J.~H.,  {Meynet} G.,   {Eldridge} J.~J.,  2022, \mn@doi [\mnras] {10.1093/mnras/stac538}, \href {https://ui.adsabs.harvard.edu/abs/2022MNRAS.512.4116F} {512, 4116}

\bibitem[\protect\citeauthoryear{{Glatzel} \& {Chernigovski}}{{Glatzel} \& {Chernigovski}}{2001}]{2001ASPC..233..227G}
{Glatzel} W.,  {Chernigovski} S.,  2001, in {de Groot} M.,  {Sterken} C.,  eds,  Astronomical Society of the Pacific Conference Series Vol. 233, P Cygni 2000: 400 Years of Progress. p.~227

\bibitem[\protect\citeauthoryear{{Glatzel} \& {Kiriakidis}}{{Glatzel} \& {Kiriakidis}}{1993}]{1993MNRAS.263..375G}
{Glatzel} W.,  {Kiriakidis} M.,  1993, \mn@doi [\mnras] {10.1093/mnras/263.2.375}, \href {https://ui.adsabs.harvard.edu/abs/1993MNRAS.263..375G} {263, 375}

\bibitem[\protect\citeauthoryear{{Glebbeek}, {Gaburov}, {de Mink}, {Pols}  \& {Portegies Zwart}}{{Glebbeek} et~al.}{2009}]{2009A&A...497..255G}
{Glebbeek} E.,  {Gaburov} E.,  {de Mink} S.~E.,  {Pols} O.~R.,   {Portegies Zwart} S.~F.,  2009, \mn@doi [\aap] {10.1051/0004-6361/200810425}, \href {https://ui.adsabs.harvard.edu/abs/2009A&A...497..255G} {497, 255}

\bibitem[\protect\citeauthoryear{{Gomez}, {Dunne}, {Eales}  \& {Edmunds}}{{Gomez} et~al.}{2006}]{2006MNRAS.372.1133G}
{Gomez} H.~L.,  {Dunne} L.,  {Eales} S.~A.,   {Edmunds} M.~G.,  2006, \mn@doi [\mnras] {10.1111/j.1365-2966.2006.10921.x}, \href {https://ui.adsabs.harvard.edu/abs/2006MNRAS.372.1133G} {372, 1133}

\bibitem[\protect\citeauthoryear{{Gormaz-Matamala}, {Cur{\'e}}, {Cidale}  \& {Venero}}{{Gormaz-Matamala} et~al.}{2019}]{2019ApJ...873..131G}
{Gormaz-Matamala} A.~C.,  {Cur{\'e}} M.,  {Cidale} L.~S.,   {Venero} R.~O.~J.,  2019, \mn@doi [\apj] {10.3847/1538-4357/ab05c4}, \href {https://ui.adsabs.harvard.edu/abs/2019ApJ...873..131G} {873, 131}

\bibitem[\protect\citeauthoryear{{Gormaz-Matamala}, {Cur{\'e}}, {Meynet}, {Cuadra}, {Groh}  \& {Murphy}}{{Gormaz-Matamala} et~al.}{2022}]{2022A&A...665A.133G}
{Gormaz-Matamala} A.~C.,  {Cur{\'e}} M.,  {Meynet} G.,  {Cuadra} J.,  {Groh} J.~H.,   {Murphy} L.~J.,  2022, \mn@doi [\aap] {10.1051/0004-6361/202243959}, \href {https://ui.adsabs.harvard.edu/abs/2022A&A...665A.133G} {665, A133}

\bibitem[\protect\citeauthoryear{{Gr{\"a}fener} \& {Vink}}{{Gr{\"a}fener} \& {Vink}}{2013}]{2013A&A...560A...6G}
{Gr{\"a}fener} G.,  {Vink} J.~S.,  2013, \mn@doi [\aap] {10.1051/0004-6361/201321914}, \href {https://ui.adsabs.harvard.edu/abs/2013A&A...560A...6G} {560, A6}

\bibitem[\protect\citeauthoryear{{Gr{\"a}fener}, {Owocki}  \& {Vink}}{{Gr{\"a}fener} et~al.}{2012}]{2012A&A...538A..40G}
{Gr{\"a}fener} G.,  {Owocki} S.~P.,   {Vink} J.~S.,  2012, \mn@doi [\aap] {10.1051/0004-6361/201117497}, \href {https://ui.adsabs.harvard.edu/abs/2012A&A...538A..40G} {538, A40}

\bibitem[\protect\citeauthoryear{{Grassitelli}, {Langer}, {Mackey}, {Gr{\"a}fener}, {Grin}, {Sander}  \& {Vink}}{{Grassitelli} et~al.}{2021}]{2021A&A...647A..99G}
{Grassitelli} L.,  {Langer} N.,  {Mackey} J.,  {Gr{\"a}fener} G.,  {Grin} N.~J.,  {Sander} A.~A.~C.,   {Vink} J.~S.,  2021, \mn@doi [\aap] {10.1051/0004-6361/202038298}, \href {https://ui.adsabs.harvard.edu/abs/2021A&A...647A..99G} {647, A99}

\bibitem[\protect\citeauthoryear{{Harpaz} \& {Soker}}{{Harpaz} \& {Soker}}{2009}]{2009NewA...14..539H}
{Harpaz} A.,  {Soker} N.,  2009, \mn@doi [\na] {10.1016/j.newast.2009.01.011}, \href {https://ui.adsabs.harvard.edu/abs/2009NewA...14..539H} {14, 539}

\bibitem[\protect\citeauthoryear{{Harpaz}, {Kovetz}  \& {Shaviv}}{{Harpaz} et~al.}{1987}]{1987ApJ...323..154H}
{Harpaz} A.,  {Kovetz} A.,   {Shaviv} G.,  1987, \mn@doi [\apj] {10.1086/165815}, \href {https://ui.adsabs.harvard.edu/abs/1987ApJ...323..154H} {323, 154}

\bibitem[\protect\citeauthoryear{{Henyey}, {Vardya}  \& {Bodenheimer}}{{Henyey} et~al.}{1965}]{1965ApJ...142..841H}
{Henyey} L.,  {Vardya} M.~S.,   {Bodenheimer} P.,  1965, \mn@doi [\apj] {10.1086/148357}, \href {https://ui.adsabs.harvard.edu/abs/1965ApJ...142..841H} {142, 841}

\bibitem[\protect\citeauthoryear{{Hubbard}}{{Hubbard}}{2007}]{2007MNRAS.374.1045H}
{Hubbard} A.,  2007, \mn@doi [\mnras] {10.1111/j.1365-2966.2006.11261.x}, \href {https://ui.adsabs.harvard.edu/abs/2007MNRAS.374.1045H} {374, 1045}

\bibitem[\protect\citeauthoryear{{Humphreys} \& {Davidson}}{{Humphreys} \& {Davidson}}{1994}]{1994PASP..106.1025H}
{Humphreys} R.~M.,  {Davidson} K.,  1994, \mn@doi [\pasp] {10.1086/133478}, \href {https://ui.adsabs.harvard.edu/abs/1994PASP..106.1025H} {106, 1025}

\bibitem[\protect\citeauthoryear{{Humphreys} \& {Martin}}{{Humphreys} \& {Martin}}{2012}]{2012ASSL..384....1H}
{Humphreys} R.~M.,  {Martin} J.~C.,  2012, in {Davidson} K.,  {Humphreys} R.~M.,  eds,  Astrophysics and Space Science Library Vol. 384, Eta Carinae and the Supernova Impostors. p.~1, \mn@doi{10.1007/978-1-4614-2275-4_1}

\bibitem[\protect\citeauthoryear{{Humphreys}, {Davidson}  \& {Smith}}{{Humphreys} et~al.}{1999}]{1999PASP..111.1124H}
{Humphreys} R.~M.,  {Davidson} K.,   {Smith} N.,  1999, \mn@doi [\pasp] {10.1086/316420}, \href {https://ui.adsabs.harvard.edu/abs/1999PASP..111.1124H} {111, 1124}

\bibitem[\protect\citeauthoryear{{Hurley}, {Pols}  \& {Tout}}{{Hurley} et~al.}{2013}]{2013ascl.soft03015H}
{Hurley} J.~R.,  {Pols} O.~R.,   {Tout} C.~A.,  2013, {SSE: Single Star Evolution}, Astrophysics Source Code Library, record ascl:1303.015 (\mn@eprint {ascl} {1303.015})

\bibitem[\protect\citeauthoryear{{Iglesias} \& {Rogers}}{{Iglesias} \& {Rogers}}{1993}]{1993ApJ...412..752I}
{Iglesias} C.~A.,  {Rogers} F.~J.,  1993, \mn@doi [\apj] {10.1086/172958}, \href {https://ui.adsabs.harvard.edu/abs/1993ApJ...412..752I} {412, 752}

\bibitem[\protect\citeauthoryear{{Iglesias} \& {Rogers}}{{Iglesias} \& {Rogers}}{1996}]{1996ApJ...464..943I}
{Iglesias} C.~A.,  {Rogers} F.~J.,  1996, \mn@doi [\apj] {10.1086/177381}, \href {https://ui.adsabs.harvard.edu/abs/1996ApJ...464..943I} {464, 943}

\bibitem[\protect\citeauthoryear{{Ishii}, {Ueno}  \& {Kato}}{{Ishii} et~al.}{1999}]{1999PASJ...51..417I}
{Ishii} M.,  {Ueno} M.,   {Kato} M.,  1999, \mn@doi [\pasj] {10.1093/pasj/51.4.417}, \href {https://ui.adsabs.harvard.edu/abs/1999PASJ...51..417I} {51, 417}

\bibitem[\protect\citeauthoryear{{Israelian} \& {de Groot}}{{Israelian} \& {de Groot}}{1999}]{1999SSRv...90..493I}
{Israelian} G.,  {de Groot} M.,  1999, \mn@doi [\ssr] {10.1023/A:1005223314464}, \href {https://ui.adsabs.harvard.edu/abs/1999SSRv...90..493I} {90, 493}

\bibitem[\protect\citeauthoryear{{Jiang}, {Cantiello}, {Bildsten}, {Quataert}  \& {Blaes}}{{Jiang} et~al.}{2015}]{2015ApJ...813...74J}
{Jiang} Y.-F.,  {Cantiello} M.,  {Bildsten} L.,  {Quataert} E.,   {Blaes} O.,  2015, \mn@doi [\apj] {10.1088/0004-637X/813/1/74}, \href {https://ui.adsabs.harvard.edu/abs/2015ApJ...813...74J} {813, 74}

\bibitem[\protect\citeauthoryear{{Johnson}}{{Johnson}}{2019}]{2019Sci...363..474J}
{Johnson} J.~A.,  2019, \mn@doi [Science] {10.1126/science.aau9540}, \href {https://ui.adsabs.harvard.edu/abs/2019Sci...363..474J} {363, 474}

\bibitem[\protect\citeauthoryear{{Justham}, {Podsiadlowski}  \& {Vink}}{{Justham} et~al.}{2014}]{2014ApJ...796..121J}
{Justham} S.,  {Podsiadlowski} P.,   {Vink} J.~S.,  2014, \mn@doi [\apj] {10.1088/0004-637X/796/2/121}, \href {https://ui.adsabs.harvard.edu/abs/2014ApJ...796..121J} {796, 121}

\bibitem[\protect\citeauthoryear{{Kandrashoff} et~al.,}{{Kandrashoff} et~al.}{2012}]{2012CBET.2976....1K}
{Kandrashoff} M.,  et~al., 2012, Central Bureau Electronic Telegrams, \href {https://ui.adsabs.harvard.edu/abs/2012CBET.2976....1K} {2976, 1}

\bibitem[\protect\citeauthoryear{{Kashi}}{{Kashi}}{2010}]{2010MNRAS.405.1924K}
{Kashi} A.,  2010, \mn@doi [\mnras] {10.1111/j.1365-2966.2010.16582.x}, \href {https://ui.adsabs.harvard.edu/abs/2010MNRAS.405.1924K} {405, 1924}

\bibitem[\protect\citeauthoryear{{Kashi}}{{Kashi}}{2019}]{2019MNRAS.486..926K}
{Kashi} A.,  2019, \mn@doi [\mnras] {10.1093/mnras/stz837}, \href {https://ui.adsabs.harvard.edu/abs/2019MNRAS.486..926K} {486, 926}

\bibitem[\protect\citeauthoryear{{Kashi} \& {Soker}}{{Kashi} \& {Soker}}{2008}]{2008MNRAS.390.1751K}
{Kashi} A.,  {Soker} N.,  2008, \mn@doi [\mnras] {10.1111/j.1365-2966.2008.13883.x}, \href {https://ui.adsabs.harvard.edu/abs/2008MNRAS.390.1751K} {390, 1751}

\bibitem[\protect\citeauthoryear{{Kashi} \& {Soker}}{{Kashi} \& {Soker}}{2010}]{2010ApJ...723..602K}
{Kashi} A.,  {Soker} N.,  2010, \mn@doi [\apj] {10.1088/0004-637X/723/1/602}, \href {https://ui.adsabs.harvard.edu/abs/2010ApJ...723..602K} {723, 602}

\bibitem[\protect\citeauthoryear{{Kashi}, {Soker}  \& {Moskovitz}}{{Kashi} et~al.}{2013}]{2013MNRAS.436.2484K}
{Kashi} A.,  {Soker} N.,   {Moskovitz} N.,  2013, \mn@doi [\mnras] {10.1093/mnras/stt1742}, \href {https://ui.adsabs.harvard.edu/abs/2013MNRAS.436.2484K} {436, 2484}

\bibitem[\protect\citeauthoryear{{Kashi}, {Davidson}  \& {Humphreys}}{{Kashi} et~al.}{2016}]{2016ApJ...817...66K}
{Kashi} A.,  {Davidson} K.,   {Humphreys} R.~M.,  2016, \mn@doi [\apj] {10.3847/0004-637X/817/1/66}, \href {https://ui.adsabs.harvard.edu/abs/2016ApJ...817...66K} {817, 66}

\bibitem[\protect\citeauthoryear{{Kashi}, {Michaelis}  \& {Kaminetsky}}{{Kashi} et~al.}{2022}]{2022MNRAS.516.3193K}
{Kashi} A.,  {Michaelis} A.,   {Kaminetsky} Y.,  2022, \mn@doi [\mnras] {10.1093/mnras/stac1912}, \href {https://ui.adsabs.harvard.edu/abs/2022MNRAS.516.3193K} {516, 3193}

\bibitem[\protect\citeauthoryear{{Langer}}{{Langer}}{1998}]{1998A&A...329..551L}
{Langer} N.,  1998, \aap, \href {https://ui.adsabs.harvard.edu/abs/1998A&A...329..551L} {329, 551}

\bibitem[\protect\citeauthoryear{{Langer}}{{Langer}}{2012}]{2012ARA&A..50..107L}
{Langer} N.,  2012, \mn@doi [\araa] {10.1146/annurev-astro-081811-125534}, \href {https://ui.adsabs.harvard.edu/abs/2012ARA&A..50..107L} {50, 107}

\bibitem[\protect\citeauthoryear{{Leitherer}}{{Leitherer}}{1997}]{1997ASPC..120...58L}
{Leitherer} C.,  1997, in {Nota} A.,  {Lamers} H.,  eds,  Astronomical Society of the Pacific Conference Series Vol. 120, Luminous Blue Variables: Massive Stars in Transition. p.~58

\bibitem[\protect\citeauthoryear{{Maeder} \& {Meynet}}{{Maeder} \& {Meynet}}{2000a}]{2000ARA&A..38..143M}
{Maeder} A.,  {Meynet} G.,  2000a, \mn@doi [\araa] {10.1146/annurev.astro.38.1.143}, \href {https://ui.adsabs.harvard.edu/abs/2000ARA&A..38..143M} {38, 143}

\bibitem[\protect\citeauthoryear{{Maeder} \& {Meynet}}{{Maeder} \& {Meynet}}{2000b}]{2000A&A...361..159M}
{Maeder} A.,  {Meynet} G.,  2000b, \mn@doi [\aap] {10.48550/arXiv.astro-ph/0006405}, \href {https://ui.adsabs.harvard.edu/abs/2000A&A...361..159M} {361, 159}

\bibitem[\protect\citeauthoryear{{Martin}, {Davidson}  \& {Koppelman}}{{Martin} et~al.}{2006}]{2006AJ....132.2717M}
{Martin} J.~C.,  {Davidson} K.,   {Koppelman} M.~D.,  2006, \mn@doi [\aj] {10.1086/508933}, \href {https://ui.adsabs.harvard.edu/abs/2006AJ....132.2717M} {132, 2717}

\bibitem[\protect\citeauthoryear{{Mauerhan} et~al.,}{{Mauerhan} et~al.}{2013}]{2013MNRAS.430.1801M}
{Mauerhan} J.~C.,  et~al., 2013, \mn@doi [\mnras] {10.1093/mnras/stt009}, \href {https://ui.adsabs.harvard.edu/abs/2013MNRAS.430.1801M} {430, 1801}

\bibitem[\protect\citeauthoryear{{Mehner}, {Davidson}, {Humphreys}, {Martin}, {Ishibashi}, {Ferland}  \& {Walborn}}{{Mehner} et~al.}{2010}]{2010ApJ...717L..22M}
{Mehner} A.,  {Davidson} K.,  {Humphreys} R.~M.,  {Martin} J.~C.,  {Ishibashi} K.,  {Ferland} G.~J.,   {Walborn} N.~R.,  2010, \mn@doi [\apjl] {10.1088/2041-8205/717/1/L22}, \href {https://ui.adsabs.harvard.edu/abs/2010ApJ...717L..22M} {717, L22}

\bibitem[\protect\citeauthoryear{{Mehner}, {Davidson}, {Humphreys}, {Ishibashi}, {Martin}, {Ruiz}  \& {Walter}}{{Mehner} et~al.}{2012}]{2012ApJ...751...73M}
{Mehner} A.,  {Davidson} K.,  {Humphreys} R.~M.,  {Ishibashi} K.,  {Martin} J.~C.,  {Ruiz} M.~T.,   {Walter} F.~M.,  2012, \mn@doi [\apj] {10.1088/0004-637X/751/1/73}, \href {https://ui.adsabs.harvard.edu/abs/2012ApJ...751...73M} {751, 73}

\bibitem[\protect\citeauthoryear{{Mehner}, {Ishibashi}, {Whitelock}, {Nagayama}, {Feast}, {van Wyk}  \& {de Wit}}{{Mehner} et~al.}{2014}]{2014A&A...564A..14M}
{Mehner} A.,  {Ishibashi} K.,  {Whitelock} P.,  {Nagayama} T.,  {Feast} M.,  {van Wyk} F.,   {de Wit} W.-J.,  2014, \mn@doi [\aap] {10.1051/0004-6361/201322729}, \href {https://ui.adsabs.harvard.edu/abs/2014A&A...564A..14M} {564, A14}

\bibitem[\protect\citeauthoryear{{Michaelis}, {Kashi}  \& {Kochiashvili}}{{Michaelis} et~al.}{2018}]{2018NewA...65...29M}
{Michaelis} A.~M.,  {Kashi} A.,   {Kochiashvili} N.,  2018, \mn@doi [\na] {10.1016/j.newast.2018.06.001}, \href {https://ui.adsabs.harvard.edu/abs/2018NewA...65...29M} {65, 29}

\bibitem[\protect\citeauthoryear{{M{\"u}ller} \& {Vink}}{{M{\"u}ller} \& {Vink}}{2014}]{2014A&A...564A..57M}
{M{\"u}ller} P.~E.,  {Vink} J.~S.,  2014, \mn@doi [\aap] {10.1051/0004-6361/201323031}, \href {https://ui.adsabs.harvard.edu/abs/2014A&A...564A..57M} {564, A57}

\bibitem[\protect\citeauthoryear{{Najarro}}{{Najarro}}{2001}]{2001ASPC..233..133N}
{Najarro} F.,  2001, in {de Groot} M.,  {Sterken} C.,  eds,  Astronomical Society of the Pacific Conference Series Vol. 233, P Cygni 2000: 400 Years of Progress. p.~133

\bibitem[\protect\citeauthoryear{{Najarro}, {Hillier}  \& {Stahl}}{{Najarro} et~al.}{1997}]{1997A&A...326.1117N}
{Najarro} F.,  {Hillier} D.~J.,   {Stahl} O.,  1997, \aap, \href {https://ui.adsabs.harvard.edu/abs/1997A&A...326.1117N} {326, 1117}

\bibitem[\protect\citeauthoryear{{Owocki}, {Gayley}  \& {Shaviv}}{{Owocki} et~al.}{2004}]{2004ApJ...616..525O}
{Owocki} S.~P.,  {Gayley} K.~G.,   {Shaviv} N.~J.,  2004, \mn@doi [\apj] {10.1086/424910}, \href {https://ui.adsabs.harvard.edu/abs/2004ApJ...616..525O} {616, 525}

\bibitem[\protect\citeauthoryear{{Papenkova} \& {Li}}{{Papenkova} \& {Li}}{2000}]{2000IAUC.7415....1P}
{Papenkova} M.,  {Li} W.~D.,  2000, \iaucirc, \href {https://ui.adsabs.harvard.edu/abs/2000IAUC.7415....1P} {7415, 1}

\bibitem[\protect\citeauthoryear{{Pastorello} et~al.,}{{Pastorello} et~al.}{2010}]{2010MNRAS.408..181P}
{Pastorello} A.,  et~al., 2010, \mn@doi [\mnras] {10.1111/j.1365-2966.2010.17142.x}, \href {https://ui.adsabs.harvard.edu/abs/2010MNRAS.408..181P} {408, 181}

\bibitem[\protect\citeauthoryear{{Pastorello} et~al.,}{{Pastorello} et~al.}{2013}]{2013ApJ...767....1P}
{Pastorello} A.,  et~al., 2013, \mn@doi [\apj] {10.1088/0004-637X/767/1/1}, \href {https://ui.adsabs.harvard.edu/abs/2013ApJ...767....1P} {767, 1}

\bibitem[\protect\citeauthoryear{{Paxton}, {Bildsten}, {Dotter}, {Herwig}, {Lesaffre}  \& {Timmes}}{{Paxton} et~al.}{2011}]{2011ApJS..192....3P}
{Paxton} B.,  {Bildsten} L.,  {Dotter} A.,  {Herwig} F.,  {Lesaffre} P.,   {Timmes} F.,  2011, \mn@doi [\apjs] {10.1088/0067-0049/192/1/3}, \href {https://ui.adsabs.harvard.edu/abs/2011ApJS..192....3P} {192, 3}

\bibitem[\protect\citeauthoryear{{Paxton} et~al.,}{{Paxton} et~al.}{2013}]{2013ApJS..208....4P}
{Paxton} B.,  et~al., 2013, \mn@doi [\apjs] {10.1088/0067-0049/208/1/4}, \href {https://ui.adsabs.harvard.edu/abs/2013ApJS..208....4P} {208, 4}

\bibitem[\protect\citeauthoryear{{Paxton} et~al.,}{{Paxton} et~al.}{2015}]{2015ApJS..220...15P}
{Paxton} B.,  et~al., 2015, \mn@doi [\apjs] {10.1088/0067-0049/220/1/15}, \href {https://ui.adsabs.harvard.edu/abs/2015ApJS..220...15P} {220, 15}

\bibitem[\protect\citeauthoryear{{Paxton} et~al.,}{{Paxton} et~al.}{2018}]{2018ApJS..234...34P}
{Paxton} B.,  et~al., 2018, \mn@doi [\apjs] {10.3847/1538-4365/aaa5a8}, \href {https://ui.adsabs.harvard.edu/abs/2018ApJS..234...34P} {234, 34}

\bibitem[\protect\citeauthoryear{{Paxton} et~al.,}{{Paxton} et~al.}{2019}]{2019ApJS..243...10P}
{Paxton} B.,  et~al., 2019, \mn@doi [\apjs] {10.3847/1538-4365/ab2241}, \href {https://ui.adsabs.harvard.edu/abs/2019ApJS..243...10P} {243, 10}

\bibitem[\protect\citeauthoryear{{Pessi}, {Prieto}  \& {Dessart}}{{Pessi} et~al.}{2023}]{2023A&A...677L...1P}
{Pessi} T.,  {Prieto} J.~L.,   {Dessart} L.,  2023, \mn@doi [\aap] {10.1051/0004-6361/202347319}, \href {https://ui.adsabs.harvard.edu/abs/2023A&A...677L...1P} {677, L1}

\bibitem[\protect\citeauthoryear{{Petrovic}, {Pols}  \& {Langer}}{{Petrovic} et~al.}{2006}]{2006A&A...450..219P}
{Petrovic} J.,  {Pols} O.,   {Langer} N.,  2006, \mn@doi [\aap] {10.1051/0004-6361:20035837}, \href {https://ui.adsabs.harvard.edu/abs/2006A&A...450..219P} {450, 219}

\bibitem[\protect\citeauthoryear{Pols, Schröder, Hurley, Tout  \& Eggleton}{Pols et~al.}{1998}]{10.1046/j.1365-8711.1998.01658.x}
Pols O.~R.,  Schröder K.-P.,  Hurley J.~R.,  Tout C.~A.,   Eggleton P.~P.,  1998, \mn@doi [\mnras] {10.1046/j.1365-8711.1998.01658.x}, 298, 525

\bibitem[\protect\citeauthoryear{{Puls}, {Springmann}  \& {Lennon}}{{Puls} et~al.}{2000}]{2000A&AS..141...23P}
{Puls} J.,  {Springmann} U.,   {Lennon} M.,  2000, \mn@doi [\aaps] {10.1051/aas:2000312}, \href {https://ui.adsabs.harvard.edu/abs/2000A&AS..141...23P} {141, 23}

\bibitem[\protect\citeauthoryear{{Quataert} \& {Shiode}}{{Quataert} \& {Shiode}}{2012}]{2012MNRAS.423L..92Q}
{Quataert} E.,  {Shiode} J.,  2012, \mn@doi [\mnras] {10.1111/j.1745-3933.2012.01264.x}, \href {https://ui.adsabs.harvard.edu/abs/2012MNRAS.423L..92Q} {423, L92}

\bibitem[\protect\citeauthoryear{{Rivet} et~al.,}{{Rivet} et~al.}{2020}]{2020MNRAS.494..218R}
{Rivet} J.~P.,  et~al., 2020, \mn@doi [\mnras] {10.1093/mnras/staa588}, \href {https://ui.adsabs.harvard.edu/abs/2020MNRAS.494..218R} {494, 218}

\bibitem[\protect\citeauthoryear{{Ro} \& {Matzner}}{{Ro} \& {Matzner}}{2016}]{2016ApJ...821..109R}
{Ro} S.,  {Matzner} C.~D.,  2016, \mn@doi [\apj] {10.3847/0004-637X/821/2/109}, \href {https://ui.adsabs.harvard.edu/abs/2016ApJ...821..109R} {821, 109}

\bibitem[\protect\citeauthoryear{{Sabhahit}, {Vink}, {Higgins}  \& {Sander}}{{Sabhahit} et~al.}{2021}]{2021MNRAS.506.4473S}
{Sabhahit} G.~N.,  {Vink} J.~S.,  {Higgins} E.~R.,   {Sander} A. A.~C.,  2021, \mn@doi [\mnras] {10.1093/mnras/stab1948}, \href {https://ui.adsabs.harvard.edu/abs/2021MNRAS.506.4473S} {506, 4473}

\bibitem[\protect\citeauthoryear{{Sabhahit}, {Vink}, {Higgins}  \& {Sander}}{{Sabhahit} et~al.}{2022}]{2022MNRAS.514.3736S}
{Sabhahit} G.~N.,  {Vink} J.~S.,  {Higgins} E.~R.,   {Sander} A. A.~C.,  2022, \mn@doi [\mnras] {10.1093/mnras/stac1410}, \href {https://ui.adsabs.harvard.edu/abs/2022MNRAS.514.3736S} {514, 3736}

\bibitem[\protect\citeauthoryear{{Sabhahit}, {Vink}, {Higgins}  \& {Sander}}{{Sabhahit} et~al.}{2023}]{2023IAUS..370..263S}
{Sabhahit} G.~N.,  {Vink} J.~S.,  {Higgins} E.~R.,   {Sander} A. A.~C.,  2023, \mn@doi [IAU Symposium] {10.1017/S1743921322003623}, \href {https://ui.adsabs.harvard.edu/abs/2023IAUS..370..263S} {370, 263}

\bibitem[\protect\citeauthoryear{{Sanyal}, {Grassitelli}, {Langer}  \& {Bestenlehner}}{{Sanyal} et~al.}{2015}]{2015A&A...580A..20S}
{Sanyal} D.,  {Grassitelli} L.,  {Langer} N.,   {Bestenlehner} J.~M.,  2015, \mn@doi [\aap] {10.1051/0004-6361/201525945}, \href {https://ui.adsabs.harvard.edu/abs/2015A&A...580A..20S} {580, A20}

\bibitem[\protect\citeauthoryear{{Shaviv}}{{Shaviv}}{2000}]{2000ApJ...532L.137S}
{Shaviv} N.~J.,  2000, \mn@doi [\apjl] {10.1086/312585}, \href {https://ui.adsabs.harvard.edu/abs/2000ApJ...532L.137S} {532, L137}

\bibitem[\protect\citeauthoryear{{Shaviv}}{{Shaviv}}{2001}]{2001ApJ...549.1093S}
{Shaviv} N.~J.,  2001, \mn@doi [\apj] {10.1086/319428}, \href {https://ui.adsabs.harvard.edu/abs/2001ApJ...549.1093S} {549, 1093}

\bibitem[\protect\citeauthoryear{{Shiber}, {Schreier}  \& {Soker}}{{Shiber} et~al.}{2016}]{2016RAA....16..117S}
{Shiber} S.,  {Schreier} R.,   {Soker} N.,  2016, \mn@doi [Research in Astronomy and Astrophysics] {10.1088/1674-4527/16/7/117}, \href {https://ui.adsabs.harvard.edu/abs/2016RAA....16..117S} {16, 117}

\bibitem[\protect\citeauthoryear{Smith}{Smith}{2011}]{smith2011explosions}
Smith N.,  2011, \mnras, 415, 2020

\bibitem[\protect\citeauthoryear{{Smith}}{{Smith}}{2013}]{2013MNRAS.429.2366S}
{Smith} N.,  2013, \mn@doi [\mnras] {10.1093/mnras/sts508}, \href {https://ui.adsabs.harvard.edu/abs/2013MNRAS.429.2366S} {429, 2366}

\bibitem[\protect\citeauthoryear{{Smith}}{{Smith}}{2017}]{2017RSPTA.37560268S}
{Smith} N.,  2017, \mn@doi [Philosophical Transactions of the Royal Society of London Series A] {10.1098/rsta.2016.0268}, \href {https://ui.adsabs.harvard.edu/abs/2017RSPTA.37560268S} {375, 20160268}

\bibitem[\protect\citeauthoryear{{Smith} \& {Arnett}}{{Smith} \& {Arnett}}{2014}]{2014ApJ...785...82S}
{Smith} N.,  {Arnett} W.~D.,  2014, \mn@doi [\apj] {10.1088/0004-637X/785/2/82}, \href {https://ui.adsabs.harvard.edu/abs/2014ApJ...785...82S} {785, 82}

\bibitem[\protect\citeauthoryear{Smith \& Owocki}{Smith \& Owocki}{2006}]{Smith_2006}
Smith N.,  Owocki S.~P.,  2006, \mn@doi [The Astrophysical Journal] {10.1086/506523}, 645, L45

\bibitem[\protect\citeauthoryear{{Smith}, {Vink}  \& {de Koter}}{{Smith} et~al.}{2004}]{2004ApJ...615..475S}
{Smith} N.,  {Vink} J.~S.,   {de Koter} A.,  2004, \mn@doi [\apj] {10.1086/424030}, \href {https://ui.adsabs.harvard.edu/abs/2004ApJ...615..475S} {615, 475}

\bibitem[\protect\citeauthoryear{{Smith} et~al.,}{{Smith} et~al.}{2007}]{2007ApJ...666.1116S}
{Smith} N.,  et~al., 2007, \mn@doi [\apj] {10.1086/519949}, \href {https://ui.adsabs.harvard.edu/abs/2007ApJ...666.1116S} {666, 1116}

\bibitem[\protect\citeauthoryear{{Smith} et~al.,}{{Smith} et~al.}{2010}]{2010AJ....139.1451S}
{Smith} N.,  et~al., 2010, \mn@doi [\aj] {10.1088/0004-6256/139/4/1451}, \href {https://ui.adsabs.harvard.edu/abs/2010AJ....139.1451S} {139, 1451}

\bibitem[\protect\citeauthoryear{{Smith}, {Li}, {Silverman}, {Ganeshalingam}  \& {Filippenko}}{{Smith} et~al.}{2011}]{2011MNRAS.415..773S}
{Smith} N.,  {Li} W.,  {Silverman} J.~M.,  {Ganeshalingam} M.,   {Filippenko} A.~V.,  2011, \mn@doi [\mnras] {10.1111/j.1365-2966.2011.18763.x}, \href {https://ui.adsabs.harvard.edu/abs/2011MNRAS.415..773S} {415, 773}

\bibitem[\protect\citeauthoryear{{Smith}, {Andrews}, {Filippenko}, {Fox}, {Mauerhan}  \& {Van Dyk}}{{Smith} et~al.}{2022}]{2022MNRAS.515...71S}
{Smith} N.,  {Andrews} J.~E.,  {Filippenko} A.~V.,  {Fox} O.~D.,  {Mauerhan} J.~C.,   {Van Dyk} S.~D.,  2022, \mn@doi [\mnras] {10.1093/mnras/stac1669}, \href {https://ui.adsabs.harvard.edu/abs/2022MNRAS.515...71S} {515, 71}

\bibitem[\protect\citeauthoryear{{Soker}}{{Soker}}{2001}]{2001MNRAS.325..584S}
{Soker} N.,  2001, \mn@doi [\mnras] {10.1046/j.1365-8711.2001.04439.x}, \href {https://ui.adsabs.harvard.edu/abs/2001MNRAS.325..584S} {325, 584}

\bibitem[\protect\citeauthoryear{{Soker}}{{Soker}}{2004}]{2004ApJ...612.1060S}
{Soker} N.,  2004, \mn@doi [\apj] {10.1086/422599}, \href {https://ui.adsabs.harvard.edu/abs/2004ApJ...612.1060S} {612, 1060}

\bibitem[\protect\citeauthoryear{{Soker} \& {Kashi}}{{Soker} \& {Kashi}}{2012}]{2012NewA...17..616S}
{Soker} N.,  {Kashi} A.,  2012, \mn@doi [\na] {10.1016/j.newast.2012.03.005}, \href {https://ui.adsabs.harvard.edu/abs/2012NewA...17..616S} {17, 616}

\bibitem[\protect\citeauthoryear{{Sundqvist}, {Puls}, {Feldmeier}  \& {Owocki}}{{Sundqvist} et~al.}{2011}]{2011A&A...528A..64S}
{Sundqvist} J.~O.,  {Puls} J.,  {Feldmeier} A.,   {Owocki} S.~P.,  2011, \mn@doi [\aap] {10.1051/0004-6361/201015771}, \href {https://ui.adsabs.harvard.edu/abs/2011A&A...528A..64S} {528, A64}

\bibitem[\protect\citeauthoryear{{Vink}}{{Vink}}{2015}]{2015ASSL..412...77V}
{Vink} J.~S.,  2015, in {Vink} J.~S.,  ed.,  Astrophysics and Space Science Library Vol. 412, Very Massive Stars in the Local Universe. p.~77 (\mn@eprint {arXiv} {1406.5357}), \mn@doi{10.1007/978-3-319-09596-7_4}

\bibitem[\protect\citeauthoryear{{Vink}}{{Vink}}{2022}]{2022ARA&A..60..203V}
{Vink} J.~S.,  2022, \mn@doi [\araa] {10.1146/annurev-astro-052920-094949}, \href {https://ui.adsabs.harvard.edu/abs/2022ARA&A..60..203V} {60, 203}

\bibitem[\protect\citeauthoryear{{Vink} \& {Gr{\"a}fener}}{{Vink} \& {Gr{\"a}fener}}{2012}]{2012ApJ...751L..34V}
{Vink} J.~S.,  {Gr{\"a}fener} G.,  2012, \mn@doi [\apjl] {10.1088/2041-8205/751/2/L34}, \href {https://ui.adsabs.harvard.edu/abs/2012ApJ...751L..34V} {751, L34}

\bibitem[\protect\citeauthoryear{{Vink}, {de Koter}  \& {Lamers}}{{Vink} et~al.}{2000}]{2000A&A...362..295V}
{Vink} J.~S.,  {de Koter} A.,   {Lamers} H.~J.~G.~L.~M.,  2000, \mn@doi [\aap] {10.48550/arXiv.astro-ph/0008183}, \href {https://ui.adsabs.harvard.edu/abs/2000A&A...362..295V} {362, 295}

\bibitem[\protect\citeauthoryear{{Vink}, {de Koter}  \& {Lamers}}{{Vink} et~al.}{2001}]{2001A&A...369..574V}
{Vink} J.~S.,  {de Koter} A.,   {Lamers} H.~J.~G.~L.~M.,  2001, \mn@doi [\aap] {10.1051/0004-6361:20010127}, \href {https://ui.adsabs.harvard.edu/abs/2001A&A...369..574V} {369, 574}

\bibitem[\protect\citeauthoryear{{Volpato}, {Marigo}, {Costa}, {Bressan}, {Trabucchi}, {Girardi}  \& {Addari}}{{Volpato} et~al.}{2024}]{2024ApJ...961...89V}
{Volpato} G.,  {Marigo} P.,  {Costa} G.,  {Bressan} A.,  {Trabucchi} M.,  {Girardi} L.,   {Addari} F.,  2024, \mn@doi [\apj] {10.3847/1538-4357/ad1185}, \href {https://ui.adsabs.harvard.edu/abs/2024ApJ...961...89V} {961, 89}

\bibitem[\protect\citeauthoryear{{Weis} \& {Bomans}}{{Weis} \& {Bomans}}{2020}]{2020Galax...8...20W}
{Weis} K.,  {Bomans} D.~J.,  2020, \mn@doi [Galaxies] {10.3390/galaxies8010020}, \href {https://ui.adsabs.harvard.edu/abs/2020Galax...8...20W} {8, 20}

\bibitem[\protect\citeauthoryear{Woosley, Heger  \& Weaver}{Woosley et~al.}{2002}]{RevModPhys.74.1015}
Woosley S.~E.,  Heger A.,   Weaver T.~A.,  2002, \mn@doi [Rev. Mod. Phys.] {10.1103/RevModPhys.74.1015}, 74, 1015

\bibitem[\protect\citeauthoryear{{Woosley}, {Blinnikov}  \& {Heger}}{{Woosley} et~al.}{2007}]{2007Natur.450..390W}
{Woosley} S.~E.,  {Blinnikov} S.,   {Heger} A.,  2007, \mn@doi [\nat] {10.1038/nature06333}, \href {https://ui.adsabs.harvard.edu/abs/2007Natur.450..390W} {450, 390}

\bibitem[\protect\citeauthoryear{{de Groot}}{{de Groot}}{1988}]{1988IrAJ...18..163D}
{de Groot} M.,  1988, Irish Astronomical Journal, \href {https://ui.adsabs.harvard.edu/abs/1988IrAJ...18..163D} {18, 163}

\bibitem[\protect\citeauthoryear{{de Jager}}{{de Jager}}{1993}]{1993SSRv...66....7D}
{de Jager} C.,  1993, \mn@doi [\ssr] {10.1007/BF00771042}, \href {https://ui.adsabs.harvard.edu/abs/1993SSRv...66....7D} {66, 7}

\bibitem[\protect\citeauthoryear{{de Koter}, {Vink}  \& {Lamers}}{{de Koter} et~al.}{1999}]{1999ASPC..192...32D}
{de Koter} A.,  {Vink} J.~S.,   {Lamers} H.~J.~G.~L.~M.,  1999, in {Hubeny} I.,  {Heap} S.,   {Cornett} R.,  eds,  Astronomical Society of the Pacific Conference Series Vol. 192, Spectrophotometric Dating of Stars and Galaxies. p.~32

\bibitem[\protect\citeauthoryear{{van Marle}, {Langer}  \& {Garc{\'\i}a-Segura}}{{van Marle} et~al.}{2004}]{2004RMxAC..22..136V}
{van Marle} A.~J.,  {Langer} N.,   {Garc{\'\i}a-Segura} G.,  2004, in {Garcia-Segura} G.,  {Tenorio-Tagle} G.,  {Franco} J.,   {Yorke} H.~W.,  eds,  Revista Mexicana de Astronomia y Astrofisica Conference Series Vol. 22, Revista Mexicana de Astronomia y Astrofisica Conference Series. pp 136--139

\bibitem[\protect\citeauthoryear{{van Marle}, {Owocki}  \& {Shaviv}}{{van Marle} et~al.}{2008}]{2008MNRAS.389.1353V}
{van Marle} A.~J.,  {Owocki} S.~P.,   {Shaviv} N.~J.,  2008, \mn@doi [\mnras] {10.1111/j.1365-2966.2008.13648.x}, \href {https://ui.adsabs.harvard.edu/abs/2008MNRAS.389.1353V} {389, 1353}

\makeatother
\end{thebibliography}
\bibliographystyle{mnras}

\end{document}